# From one to three dimensions – corrugated $^1_\infty$[NiGe] ribbons as building block in alkaline-earth metal *Ae/*Ni/Ge phases. Crystal structure and chemical bonding in *Ae*NiGe (*Ae* = Mg, Sr, Ba)

**Viktor Hlukhyy\*, Lisa Siggelkow, Thomas F. Fässler**

Departement of Chemistry, Technische Universität München, Lichtenbergstr. 4, 85747 Garching , Germany

**Abstract.**

New equiatomic nickel germanides MgNiGe, SrNiGe, and BaNiGe have been synthesized from the elements in sealed tantalum tubes using a high-frequency furnace. The compounds were investigated by X-ray diffraction both on powders and single crystals. MgNiGe crystallizes with TiNiSi structure type, space group *Pnma*, *Z* = 4, *a* = 6.4742(2) Å, *b* = 4.0716(1) Å, *c* = 6.9426(2) Å, w$R_2$ = 0.033, 305 $F^2$ values, 20 variable parameters. SrNiGe and BaNiGe are isotypic and crystallize with anti-SnFCl structure type (*Z* = 4, *Pnma*) with *a* = 5.727(1) Å, *b* = 4.174(1) Å, *c* = 11.400(3) Å, w$R_2$ = 0.078, 354 $F^2$ values, 20 variable parameters for SrNiGe and *a* = 5.969(4) Å, *b* = 4.195(1) Å, *c* = 11.993(5) Å, w$R_2$ = 0.048, 393 $F^2$ values, 20 variable parameters for BaNiGe. The increase of the cation size leads to a reduction of the dimensionality of [NiGe] polyanions. In the MgNiGe structure the nickel and germanium atoms build a three-dimensional $^3_\infty$[NiGe] network with magnesium atoms in the channels. In contrast, in Sr(Ba)NiGe one-dimensional $^1_\infty$[NiGe] ribbons, running parallel to the *b*-axis, are separated by strontium/barium atoms. The crystal chemistry and chemical bonding in equiatomic alkaline-earth nickel germanides $Ae$NiGe ($Ae$ = Mg, Ca, Sr, Ba) is discussed. The experimental results are reconciled with electronic structure calculations performed using the tight-binding linear muffin-tin orbital (TB-LMTO-ASA) method.

**Keywords:** Intermetallic Compounds, One-Dimensional Polyanion, Ribbon, Germanide

**Address for correspondence:**

* Dr. Viktor Hlukhyy, Department of Chemistry, Technische Universität München, Lichtenbergstr. 4, D-85747 Garching Germany, E-mail: viktor.hlukhyy@lrz.tu-muenchen.de

## Introduction

The Zintl-Klemm concept provides a universal way to describe the relationship between crystal structure and chemical bonding in a large variety of intermetallic compounds containing electropositive metals such as alkali ($A$) or alkaline-earth ($Ae$) metals and a p-block (semi)metal ($E$).[1] Localized chemical bonding pictures predict the connectivity of the polyanions consisting of p-block atoms $E$ according to the valence rules. Similar to the polyanions of $[E_n]^{m-}$ in Zintl phases, in ternary polar intermetallic phases containing additional transition metals ($T$) specific atom arrangments $[T_pE_n]^{m-}$ occur recurrently. Even though metallic properties are observed in the corresponding band structures, a *formal* electron transfer from the $A$ or $Ae$ atoms leads to polyanionic units $[T_pE_n]^{m-}$ in the case of ternary $A_mT_pE_n$ and $Ae_{m/2}T_pE_n$ intermetallic compounds.

Unlike Zintl phases, polar intermetallic compounds show more delocalized chemical bonding in their polyanionic networks compared to the classical 2c-2e chemical bonds. Hovever, like to Zintl phases in the polar intermetallic phases $Ae_{m/2}$Ni$_p$Ge$_n$ the increasing of *Ae* amounts leads to a reduction of the dimensionality of the polyanionic $[Ni_pGe_n]^{m-}$ structures. So far, the $[Ni_pGe_n]^{m-}$ substructures in alkaline-earth nickel germanides occur mainly as three-dimensional polyanionic networks (in $MgNi_6Ge_6$, $Mg_6Ni_{16}Ge_7$, $CaNi_2Ge_2$, $CaNi_5Ge_3$, $CaNiGe_3$, $CaNiGe_2$, $Ca_7Ni_{49}Ge_{22}$, $Ca_{15}Ni_{68}Ge_{37}$, $Ca_5Ni_{17}Ge_8$, $SrNi_2Ge_2$, $SrNiGe_2$, $SrNiGe_3$, $Ba_8Ni_{3.5}Ge_{42.1}$) [2-8] or two-dimensional Ni-Ge polyanionic layers (in CaNiGe, $Ca_2Ni_3Ge_2$, $Ca_4Ni_4Ge_3$, $SrNi_3Ge_2$, $SrNi_2Ge$, $BaNi_2Ge$, $Ba_2Ni_5Ge_4$, $BaNi_2Ge_2$),[9-14] where a reduction of the dimensions of the $[Ni_pGe_n]^{2m-}$ substructures is observed with increasing of alkaline-earth metal content.

Besides the *Ae* content in ternary intermetallic phases $Ae_{m/2}Ni_pGe_n$, we study the influence of the size of the *Ae* atoms on the form and dimensionality of Ni-Ge polyanions. In this context we studied a serie of equiatomic compounds *Ae*NiGe (*Ae* = Mg, Ca, Sr, and Ba). Here we report on the new compounds SrNiGe and BaNiGe whose crystal structures posses unique one-dimensional $^1_\infty$[NiGe] ribbons as well as the new compound MgNiGe in which a three-dimensional $^3_\infty$[NiGe] network with exlusively four-bonded Ni and Ge appears as polyanionic substructure. Calculations of the electronic structure, including the analysis of the band structure, the COHP and the ELF, which are indicative for chemical bonding complement the experimental findings.

## Experimental Section

**Synthesis.** Starting materials for the synthesis of MgNiGe, SrNiGe, and BaNiGe were ingots of the magnesium (ChemPur), strontium (ChemPur), barium (ChemPur), nickel wire ($\varnothing$ 1 mm, Johnson-Matthey), and germanium pieces (ChemPur), all with stated purities better than 99.5%. Pieces of the alkaline-earth metals, pieces of nickel wire, and pieces of germanium were mixed in a 1:1:1 atomic ratio. The mixtures for samples were subsequently sealed in tantalum tubes under argon atmosphere (Mini Arc Melting System, MAM-1, Johanna Otto GmbH, placed in an argon filled glovebox). The crucibles were placed in a water-cooled sample chamber of an induction furnace (Hüttinger Elektronik, Freiburg, Typ TIG 2.5/300), heated under flowing argon up to approximately 950 °C and kept at that temperature for 30 minutes. The reactions between the elements were visible through a slight flash of light. After the melting procedure the samples were cooled within one hour to approximately 600 °C and held at that temperature for another one hour, finally quenched to room temperature by switching off the furnace. The temperature above 900 °C was controlled through a Sensor Therm Metis MS09 pyrometer with an accuracy of ±10 K. After cooling to room temperature, the gray samples could easily be separated from the tantalum crucibles. No reactions of the samples with the crucible material could be detected. MgNiGe sample is stable in moist air as fine-grained powders, whereas the powdered samples of SrNiGe and BaNiGe are stable only for few days. Good quality irregular form crystals in case of MgNiGe and needle-like single crystals in case of SrNiGe and BaNiGe with metallic lustre were isolated from the crushed samples prepared in the induction furnace. SrNiGe and BaNiGe can by easily reproduced by arc melting technique (Mini Arc Melting System, MAM-1, Johanna Otto GmbH, placed in an argon filled glovebox), whereas the syntheses of MgNiGe can only by carried out successfuly in welded Ta/Nb ampoules due to the high vapour pressure of Mg.

**X-Ray Investigations.** The purity of the sample was checked using a STOE STADI P powder diffractometer with with Ge monochromatized CuK$_\alpha$ radiation ($\lambda$ = 1.54056 Å). The ortorhombic lattice parameters (Table 1) were obtained from least-squares fit of the powder data of MgNiGe, SrNiGe, and BaNiGe. The correct indexing of the patterns was ensured through intensity calculations taking the atomic positions from the structure refinements obtained from single crystal X-ray diffraction measurement. In all cases the lattice parameters determined from powder patterns and from single crystal data agreed well. The Rietveld refinement with the Fullprof Suite [15] showed that except for the main phase the MgNiGe contained about 4% of $Mg_2Ge$ [16] and 7% $MgNi_6Ge_6$, [17] whereas SrNiGe and BaNiGe were single-phase ones within the accuracy of the powder diffraction method (Supporting Information, Figures S1–S3).

Single crystal of MgNiGe was measured at room temperature on an Oxford-Xcalibur3 diffractometer (CCD area detector) with graphite monochromatized $MoK_\alpha$ ($\lambda = 0.71073$ Å) radiation. Single crystal data sets of SrNiGe and BaNiGe were collected in the $\phi$-scan mode using a Stoe IPDS-IIT imaging plate detector diffractometer (Mo-$K_\alpha$ radiation, $\lambda = 0.71073$ Å, graphite monochromator) at room temperature. Numerical absorption corrections with optimized crystal shapes using the X-SHAPE and X-RED programs (X–Shape/X–Red) [18, 19] were performed for SrNiGe and BaNiGe, whereas for MgNiGe the absorption was corrected empirically.[20] The all structures were solved using direct methods (SHELXS [21]) and refined by full-matrix least squares using SHELXL.[22] Direct methods provided the positions of atoms in the orthorhombic primitive space group *Pnma*. The occupation factors of all atoms did not deviate from unity. All final cycles included anisotropic displacement parameters and revealed no significant residual peaks. All relevant crystallographic data for the data collection and evaluation are listed in Table 1. Final atomic positions with equivalent displacement parameters are given in Table 2 and selected bond lengths in Table 3.

After data collection the single crystals were analyzed by EDX measurement with a JEOL SEM 5900LV scanning electron microscope equipped with an Oxford Instruments INCA energy dispersive X-ray microanalysis system. No impurity elements heavier than sodium have been observed. The analysis of well-shaped single crystals has revealed its compositions (in atomic percentages): Mg 27(3), Ni 36(4), and Ge 37(5) for MgNiGe; Sr 29(3), Ni 36(5), and Ge 35(5) for SrNiGe; and Ba 36(4), Ni 33(4), and Ge 31(5) for BaNiGe.

The values within the standard deviations are in good agreement with the ideal equiatomic compositions.

**Magnetic Susceptibility Measurements.** DC magnetization data were collected using a Quantum Design MPMS XL5 superconducting quantum interference device (SQUID). The temperature-dependent data were obtained by measurement of the magnetization from 1.8 to 300 K in an applied magnetic field of 5 kOe and 1 kOe for BaNiGe and SrNiGe, respectively, by using the powdered samples held in a straw.

**Electronic Structure Calculations.** The electronic structure was investigated by means of the *ab initio* linear muffin-tin orbital (LMTO) method in the atomic sphere approximation (ASA), using the tight-binding (TB) program TB-LMTO-ASA[23]. The basis set of short-ranged[24] atom-centered TB-LMTOs contained s-p valence functions for Mg, s-f valence functions for Sr, Ba, and s-d valence functions for Ni and Ge. Mg 3p; Sr 5p, 4f; Ba 6p; and Ge 3d orbitals were included using a downfolding technique.[25] To achieve space filling within the atomic sphere approximation, interstitial spheres are introduced to avoid too large overlap of the atom-centered spheres. The empty sphere positions and radii were calculated using an automatic procedure. We did not allow an overlap of more than 16% for any two atom-centered spheres. In the fatband analysis the atomic orbital character is represented as a function of the band width. For analysing the band structure of MgNiGe, SrNiGe and BaNiGe the following *k*-path has been chosen: Γ = (0,0,0), Z = (0,0,½), T = (0,½,½), Y= (0,½,0), Γ = (0,0,0), X = (½, 0, 0), S = (½,½,0), R = (½,½,½) and U = (½,0,½). The total and partial Density-Of-States (DOS) were computed and studied. The Crystal Orbital Hamilton Populations (COHPs)[26] were employed for analysis of the chemical bonding. From COHP analyses, the contribution of the covalent part of a particular interaction to the total bonding energy of the crystal can be obtained. All COHP curves are presented here in the following format: positive region indicate bonding and negative regions show antibonding interactions. The Fermi level $E_F$ was set as a reference point at 0 eV.

## Results and Discussion

**Description of the Crystal Structures.** *MgNiGe* crystallizes with the well known TiNiSi-type structure (space group *Pnma*, such as many other equiatomic germanides and silicides [27, 28]). The structure derives from the well known aristotype $AlB_2$ via an ordering of the Ni and Ge atoms on the graphite-type boron layer of $AlB_2$ in analogy to boron nitride. The $3^6$ nets parallel to the *bc* plane are strongly puckered and heteroatomic bonds are formed in between the layers.

The resulting three-dimensional NiGe structure with alternating Ni and Ge atoms consists of four-, six-, and eight-membered Ni-Ge polygons. The largest eight membered polygons form channels along the *b* axis, which are occupied with Mg atoms (Figure 1a). The Ni–Ge distances range from 2.391(1) to 2.436(1) Å and compare well to the sum of Pauling's single bond radii of 2.44 Å.[29] There are no close Ni–Ni and Ge–Ge contacts within $^3_\infty[NiGe]$ network. All atoms of the network are four connected with strongle distorted thetrahedral coordination to neighbouring atoms. The Ni-Ge-Ni and Ge-Ni-Ge bond angles range from 75.89(2)° to 121.88(2)° and from 100.38(2)° to 116.74(2)°, respectively. In the case of Ge the distortion leads to trigonal pyramids of neighbouring Ni atoms, the centering Ge atoms being situated close to the basal plane. Each Mg in MgNiGe atom has twelve nearest Ni/Ge neighbors within the distances range of 2.737(1) Å to 3.283(1) Å, the coordination polyhedron being a distorted hexagonal prism $Ni_6Ge_6$.

Due to the puckering of the [NiGe] layers, the orthorhombic distortions of the [NiGe] polyanion and the formation of relatively short Ni-Ni bonds, the [NiGe] ribbons consisting of a pleated sheet structure of rhombic $Ni_2Ge_2$ units emerge (Figure 2a). In these rhombic $Ni_2Ge_2$ units the more electronegative Ge atoms occupy the energetically favored position in order to maximize the Ge–Ge distance (minimize the repulsion): the Ge–Ge distance of 3.779(1) Å is significantly longer than the Ni–Ni distance of 2.946(1) Å. Such trend of the tilting of the $T_2X_2$ units was recently discussed for different representatives of the TiNiSi structure type, which proved to be flexible against the exchange of atoms.[30, 31] An organic analogue to isolated $^1_\infty[Ni_2Ge_2]$ ribbons is cyclobutane ladder polymer. Ladder polysilanes -$(Si_2R_2)_n$- were theoretical studied using density functional theory.[32]

In order to compare the MgNiGe structure with the forthcoming $Ae$NiGe structures we emphasize in Figure 1b and 1c the two-atom wide $^1_\infty[Ni_2Ge_2]$ zig-zag ribbon with $d$(Ni-Ge) = 2.391(1) and 2.401(1) Å . The rombic $Ni_2Ge_2$ units are interconnected along the $b$ direction to a pleated sheet structure ($^1_\infty[Ni_2Ge_2]$ ribbons) with a dihedral angle α between the rhombic $Ni_2Ge_2$ units of 122.80(2)° (Figure 1a and 1b, Table 4, Sheme 1). These ribbons are further interconnected to a three-dimensional network by Ni-Ge bonds which are slightly longer (2.436(1) Å).

*SrNiGe* and *BaNiGe* are the first representatives of intermetallic compounds that crystallize with anti-SnClF type structure (space group *Pnma*). The crystal structure consists of unique one-dimensional two-atom wide zig-zag $^1_\infty[NiGe]$ ribbons running along the $b$ axis and are separated by Sr and Ba atoms, respectively (Figure 1d). These $^1_\infty[NiGe]$ ribbons are similar to those described for MgNiGe. They consist of rhombic $Ni_2Ge_2$ units, which are interconnected along $b$ via Ni-Ge bonds, such that a pleated sheet structure results.

The observed Ni–Ge distances within $^1_\infty[NiGe]$ ribbons (2×2.307(1) Å and 2.387(1) Å for SrNiGe as well as 2×2.317(1) Å and 2.372(1) Å for BaNiGe) are shorter than other known distances in $Ae$/Ni/Ge ($Ae$ = Mg, Ca, Sr and Ba) compounds [2-7, 9-14] and are also shorter than the Pauling single bond distances (2.44 Å). However, they are comparable to those in binary nickel germanides, for example in equiatomic NiGe (from 2.330 to 2.487 Å) [24].

The rhombic distortion from a $Ni_2Ge_2$ square is more expressed in the Sr and Ba containing compounds compared to the intermetallic phase MgNiGe as indicated by the deviation of the Ni-Ge-Ni angle from 90° (Mg: 75.88(2)°, Sr 66.83(5)°, Ba 66.91(2)°). As a result additional Ni–Ni bonds with a length of 2.586(1) Å for both SrNiGe and BaNiGe (in contrast to $d$(Ni-Ni) = 2.946(1) Å for MgNiGe) form and zigzag Ni-chains within the polymeric $^1_\infty$[NiGe] ribbons appear (Figure 1e). In comparison to *fcc* nickel (2.49 Å),[29] the Ni-Ni distances are elongated. Consequently, the zig-zag ribbon (pleated sheet structure) is flattend in comparison to MgNiGe. Thus, a dihedral angle of 159.30(7)° and 159.60(5)° results for SrNiGe and BaNiGe, respectively (see Table 4).

The Ni-Ge and Ni-Ni distances within the $^1_\infty$[NiGe] ribbon are only slightly affected by the nature of the *Ae* atoms. This is reflected by the the values of the unit cell parameters going from SrNiGe to BaNiGe: while the parameters *a* and *c*, which determine the inter-ribbon contacts, are increased by 4.2% and 5.2%, respectively, the parameter *b* that measures the length of $^1_\infty$[NiGe] ribbons changes only slightly (0.5%). This anisotropic behavior is a strong indication that the chemical Ni-Ge bonds play a dominant role. This strong interaction within $^1_\infty$[NiGe] ribbons underlines their polyanionic character.

Examining the coordination of the nickel atoms within polyanion a $Ni_2\square Ge_3$ hexagon with one vacant Ni position is found in compare to those of recently published $Sr(Ba)Ni_2Ge$ (Figure 2e,f),[11] i.e. each nickel atom has three germanium and two nickel neighbors in a almost planar coordination. Furthermore, each nickel atom has five nearest Sr(Ba) neighbours, two above and three below this defect hexagon. Each Ge atom is coordinated by three Ni and seven Sr(Ba) atoms. Sr(Ba) has twelve atoms in their coordination shell (five Ni and seven Ge). There are no close Ge-Ge contacts within Sr(Ba)NiGe structure.

**Structural Relationships.** Comparative studies of the structure of intermetallic germanides $Ae$/Ni/Ge reveal that in many cases $^1_\infty$[NiGe] ribbons are the building block of [Ni$_p$Ge$_n$]$^{m-}$ polyanionic substructures. The three title structures nicely flank the known CaNiGe structure [9], in which a two-dimensional $^2_\infty$[NiGe] substructure appears. This PbO-type $^2_\infty$[NiGe] layer can be interpreted as parallel aligned $^1_\infty$[NiGe] ribbons which are interconnected perpendicularly to the direction of the ribbon. As the inter- and intraribbon Ni-Ge distances are equal, this remains a topological model. However, it allows to uncover structural relationships. Thus, all three structures shown in Figure 2a-c can be traced back to corrugated $^1_\infty$[NiGe] ribbons. These ribbons compose a three-dimensional framework in MgNiGe, two-dimensional puckered layer in the tetragonal CaNiGe structure and are as isolated one-dimensional ribbons in SrNiGe and BaNiGe. In the same order, the pleated sheet structure is flattened (the dihedral angles rise from 122.80(2)° in MgNiGe, to 126.20(1)° in CaNiGe to 159.30(7)° and 159.60(5)° in SrNiGe and BaNiGe, respectively) and the intra-ribbon Ni-Ni bond remains almost the same going from MgNiGe to CaNiGe (2.947(1) Å and 2.965(1) Å, respectively) and is drastically shortened in SrNiGe and BaNiGe (2.586(2) Å and 2.588(2) Å). Thus, the decrease of the dimensionality leads to a flattening of the pleated sheet structure and to a shortening of the Ni-Ni bond.

In analogy, the coordination polyhedra of nickel and germanium atoms change from strongly distorted tetrahedra in MgNiGe (angle (Ge-Ni-Ge) = 100.37(2)° to 116.74(2)°; angle (Ni-Ge-Ni) = 75.89(2)° to 121.54(2)°). In CaNiGe the Ni atoms are tetrahedrally coordinated (angle (Ge-Ni-Ge) = 104.9° – 119.0°) and Ge atoms have an umbrella type coordination (angle (Ni-Ge-Ni) = 75°). In SrNiGe and BaNiGe the coordination numbers of Ni and Ge within the NiGe ribbon is reduced to three, however the Ni atoms form additional Ni-Ni bonds along the diagonal of the rhomb (shown as grey lines in Figure 2c).

Interestingly, the Ni-Ge substructure of other recently described nickel germanides can also be traced back to corrugated $^1_\infty$[NiGe] ribbons. For example, the polar intermetallic compound $Ca_4Ni_4Ge_3$ contains $^2_\infty$[Ni$_4$Ge$_3$] layers.[11] These can be described as being built up of the $^1_\infty$[NiGe] ribbons (with a dihedral angle of 151.90°), which are interconnected by Ni–Ge–Ni bridges, as shown in Figure 2d. The $^2_\infty$[Ni$_2$Ge] layers of $SrNi_2Ge$ and $BaNi_2Ge$ are composed by corrugated Ge-centered Ni hexagons with chair and boat conformation, respectively, like two-dimensional graphane nets.[33] However, the layer can also be described as being built up of $^1_\infty$[NiGe] ribbons that are interconnected by a parallel oriented Ni zig-zag chains (Figure 2e and 2f, respectively).[11-12] The polar intermetallic compound $SrNi_3Ge_2$ contains $^2_\infty$[Ni$_3$Ge$_2$] layers of condensed hexagonal Ni-centered prisms. These can also be described being built up of $^1_\infty$[NiGe] ribbons that correspond now to the prism heights. The parallel oriented ribbons are connected by Ni atoms (Figure 2g).[12]

In all these structures, containing $^1_\infty$[NiGe] ribbons as building blocks, one interesting relation is observed: by flattening the ribbons corrugation (increasing of the dihedral angle between fused $Ni_2Ge_2$ rombs) the additional Ni–Ni bonds within rombs appears despite almost unchanged Ni–Ge distances within the ribbon (see Table 4).

Further, the structures of $Ae$Ni$_2$Ge$_2$ ($Ae$ = Ca, Sr, Ba)[2, 14] contain [NiGe] layers similar to those of CaNiGe. Due to the lower amount of alkaline-earth metal a different stacking of the layers results. Consequently, short interlayer Ge-Ge bonds are formed in CaNi$_2$Ge$_2$ ($d$(Ge-Ge) = 2.61 Å). These bonds are widened in SrNi$_2$Ge$_2$ and HT-BaNi$_2$Ge$_2$ ($d$(Ge-Ge) = 2.83 Å and $d$(Ge-Ge) = 3.64 Å, respectively). The [NiGe] ribbons can also be found in structures of other recently published Ni-rich germanides CaNi$_5$Ge$_3$, Ca$_{15}$Ni$_{68}$Ge$_{37}$ and Ca$_7$Ni$_{49}$Ge$_{22}$.[3] These ribbons are highly condensed and represent sections of the distorted Ni$_3$Ge structure.

In a more general view corrugated $^1_\infty$[$T$Ge] ribbons appear as part of two- or three-dimensional networks in various transition metal ($T$) germanides. For example, corrugated and parallel oriented $^1_\infty[T\mathrm{Ge}]$ ribbons ($T$ – transition metal) can also be emphasized in $\mathrm{Sm_3Co_2Ge_4}$ in which $^1_\infty[\mathrm{CoGe}]$ ribbons are interconnected via parallel oriented $^1_\infty[\mathrm{Ge}]$ zig-zag chains resulting in a two-dimensional $[\mathrm{CoGe_2}]$ layer (Supporting Information, Figure S4d).[34] In $Re_4\mathrm{Ni_2InGe_4}$ structures ($Re$ = Dy, Ho, Er, Tm) the connection between paralell oriented $^1_\infty[\mathrm{NiGe}]$ ribbons is achieved by $^1_\infty[\mathrm{Ge_2In}]$ zig-zag chains (Supporting Information, Figure S4a).[35] In monoclinic YbFeGe $^1_\infty[\mathrm{FeGe}]$ ribbons occur. These ribbons are connected by Ge–Ge bonds forming two-dimensional $^2_\infty[\mathrm{FeGe}]$ layers (Supporting Information, Figure S4b).[36] In $\mathrm{Lu_3Ir_2Ge_3}$ and $\mathrm{Yb_2IrGe_2}$ such layers of Ge–Ge connected $^1_\infty[T\mathrm{Ge}]$ ribbons are further connected by bridging Ge atoms and $\mathrm{Ge_2}$ dumbbells, respectively, into three-dimensional $T$– Ge frameworks (Supporting Information, Figure S4c, e).[37]

One-dimensional chains composed solely of nickel and germanium atoms are further found as building blocks in other rare-earth nickel germanides. Even if the equimolar composition of the Ni-Ge polymers is retained, other topologies can result: Two different types of chains as well as isolated Ge atoms are found in $\mathrm{La_{11}Ni_4Ge_6}$.[27] Ni-Ni edge sharing $\mathrm{Ni_4Ge_2}$ hexagons form planar $^1_\infty[\mathrm{Ni_2Ge_2}]$ bands, whereas the second polymeric unit forms planar polyacetylene-type $^1_\infty[\mathrm{Ni}]$ zig-zag chains with exo-bonded Ge atoms (Figure 3a–c). Finally, $\mathrm{La_3NiGe_2}$ possesses a similar heteroatomic planar polyacetylene-type $^1_\infty[\mathrm{NiGe}]$ zig-zag chain with syntactic oriented Ge atoms bound to the Ni atoms (Figure 3d–e).[27] The formal addition of a further Ni atom as shown in Figure 3f leads to the topology of the ribbons of the title compounds. Thus, the polymer in $\mathrm{La_3NiGe_2}$ can be seen as defective variant of the $^1_\infty[\mathrm{NiGe}]$ ribbons of Ba(Sr)NiGe.

**Group-subgroup relationships in MgNiGe, CaNiGe and Sr(Ba)NiGe.** Rather interesting similarities between equiatomic alkaline-earth nickel germanides and their lead or tin halogenide prototypes exist. Although MgNiGe (TiNiSi structure type, $\mathrm{PbCl_2}$ binary prototype) and Sr(Ba)NiGe (anti-SnFCl structure type) structures appear with the same space group *Pnma* and the respective atoms occupy the same Wyckoff sites, it exists an isopointal rather than an isotypic relationship, due to the significant differences in chemical bonding. The recently published CaNiGe [9] crystallizes in CeFeSi (or anti-PbFCl) structure type (space group *P4/nmm*).

The halogenide $PbCl_2$ (binary prototype of TiNiSi structure in which MgNiGe crystallizes) [38] topologically contains a three-dimensional $^{3}_{\infty}[PbCl]^{+}$ network (Supporting Information, Figure S5a) similarly to the $[NiGe]^{2-}$ polyanionic network of MgNiGe. In contrast, in the isoelectronic PbFCl structure type two-dimensional $^{2}_{\infty}[PbF]^{+}$ layers with PbO-like topology are present,[27] corresponding to the $[NiGe]^{2-}$ polyanionic layers in CaNiGe (Supporting Information, Figure S5b). Furthermore, in the SnFCl type structure (Supporting Information, Figure S5c),[39] which can be considered as prototype of Sr(Ba)NiGe, topologically the one-dimensional $^{1}_{\infty}[SnF]^{+}$ ribbons can be emphasized, in analogy to the $[NiGe]^{2-}$ polyanionic ribbons Sr(Ba)NiGe. Most probably, the size factor plays a major role on the structure formation of halogenides as well, however, in contrast to intermetallics – in the atomic positions of the "polycationic" units.

In the following we will discuss the structural relations between CaNiGe (PbFCl-type, *P4/nmm*) and Sr(Ba)NiGe (SnFCl-type, *Pnma*) or MgNiGe (TiNiSi- or $PbCl_2$-type, *Pnma*) in more detail on the basis of the group-subgroup scheme, a compact graphical representation which has been introduced by *Müller* and *Bärnighausen*.[40, 41] As mentioned above, MgNiGe and Sr(Ba)NiGe crystallize in the same space group *Pnma* and are isopointal. The crystal structures of CaNiGe (*P4/nmm*) and Mg(Sr,Ba)NiGe (*Pnma*) are not directly related via a group-subgroup scheme, however, are linked in a interesting way by the orthorhombic space group *Pmmn* (FeOCl structure type).[42]

The symmetry reduction from tetragonal symmetry (CaNiGe) to the orthorhombic symmetry (MgNiGe and Sr(Ba)NiGe) allows a strong uncoupling of coordinates of the atoms within the square $^2_\infty$[NiGe] layers of CaNiGe along with a puckering of the atomic layers (Figure 4).

Starting with CaNiGe and moving one half of the nickel atoms above and the other half below the plane at z = 0 at the same distance leads to the crystal structure of FeOCl (Figure 4b; Ca atoms are replaced by Cl, Ni by O and Ge by Fe). Due to this atom movement the symmetry $P4/n2_1/m2/m$ is reduced by a *translationengleiche* reduction of index 2 (t2) leading to the orthorhombic space group $P2_1/m2_1/m2/n$.

Doubling the unit cell of FeOCl along the *c* axis and lowering the symmetry of the space group by a *klassengleiche* transition of index 2 (k2) to $P2_1/n2_1/m2_1/a$ leads to free parameters *x* and *z* for all atoms in MgNiGe and (Sr/Ba)NiGe. Thus, uncoupling the positions of the atoms of the original square layer leads to 1D ribbons, which are isolated in case of Sr(Ba)NiGe or linked via Ni-Ge bonds to a 3D network in case of MgNiGe.

**Magnetic properties.** Thermal dependencies of the molar and inverse magnetic susceptibilities $\chi_{mol}$ for both of SrNiGe and BaNiGe in the range 1.8–300 K are presented in Figure S6 (Supporting Information). Obtained raw magnetization data were converted into molar magnetic susceptibilities ($\chi_{mol}$) and subsequently corrected for the holder and for the diamagnetic contribution of the core electrons. Neither indications of superconductivity nor long range magnetic order were observed down to a temperature of 1.8 K for BaNiGe and SrNiGe. Above 50 K the susceptibility of BaNiGe and SrNiGe has been found nearly field-independent and following Curie-Weiss behavior, where local moments are supposed to arise from unpaired d-electrons of nickel. Fitting of the obtained magnetic susceptibilities (above 50 K for BaNiGe and above 100 K for SrNiGe) using the Curie-Weiss law results in an effective magnetic moment $\mu_{eff} = 1.40(1)$ $\mu_B$ and $\mu_{eff} = 2.00(5)$ $\mu_B$ per Ni atom in BaNiGe and SrNiGe, respectively. Compared to the theoretical value for $Ni^{2+}$ (2.83 µB) the obtained magnetic moments have been found rather small, which along with the extremely high negative $\theta_c$ might indicate a presence of short-range magnetic ordered clusters, typical for low-dimensional magnets. On the other hand the deviations from Curie-Weiss behavior below 50 K may be attributed to a minor trace of ferromagnetic impurities (most likely nickel), although our powder diffractograms showed single phase SrNiGe and BaNiGe samples. No indications of superconductivity for MgNiGe were observed down to a temperature of 1.8 K.

**Electronic Structure Calculations.** TB-LMTO-ASA electronic structure calculations were carried out for MgNiGe, SrNiGe, and BaNiGe. In order to examine the electronic structure in detail and to compare the chemical bonding, the partial DOS, band structures including fatbands, COHP and ELF have been calculated.

In all four discussed equiatomic compounds MgNiGe, CaNiGe, SrNiGe and BaNiGe the alkaline earth metal atoms are the most eletropositive components (Pauling's electronegativities are: 1.3 for Mg, 1.0 for Ca, 1.0 for Sr, 1.0 for Ba, 1.9 for Ni, and 2.0 for Ge). The have largely transferred their valence electrons to the [NiGe] network. To a first approximation, the formula may be written as $Ae^{2+}[NiGe]^{2-}$, emphasizing the covalent Ni–Ge bonding within the polyanion. Although the description by a polyanionic network seems adequate at first sight, some words of caution seem to be appropriate, since significant bonding interactions also occur between the alkaline earth metal and the [NiGe] network. Especially the Mg–Ni and Mg–Ge interactions need to be considered (see shortest distances in Table 3; d(Mg–Ni) = 2.737(1) Å, d(Mg–Ge) = 2.800(1) Å).

The calculated total and partial density of states (DOS) curves for MgNiGe, CaNiGe, SrNiGe and BaNiGe are shown in Figure 5. For all compounds a non-zero value of the DOS is observed at $E_F$, thus indicating metallic conductivity. The total DOS curves of all four compounds can be divided in three parts. The first one (below $-8$ eV) has mainly Ge(s) character with minor contributions of Mg(s), Mg(p) or $Ae$(d) ($Ae$: Ca, Sr, Ba) as well as Ni(d) orbitals. The second one (approximately from $-6.5$ eV to $-4$ eV for MgNiGe and from $-5$ eV to −3 eV for $Ae$NiGe) is dominated by the Ge(p) orbitals as well as the Mg(s) and Mg(p). In contrast, the third section of the total DOS curve (above −4 eV for MgNiGe and above −3 eV for $Ae$NiGe) is mainly dominated by the Ni(d) orbitals.

Analysing the first part of the total DOS curves (below −8 eV) for CaNiGe, SrNiGe and BaNiGe sharp peak is observed (Figures 5b−d), whereas for MgNiGe the DOS in the Ge(s) orbital region is rather flat (Figure 5a). This is in good agreement with the coordination numbers of the respective Ge atoms: in SrNiGe and BaNiGe the Ge atoms have three nearest neighbours of Ni, in CaNiGe the Ge atoms are situated on top of a square pyramid GeNi$_4$ and in MgNiGe the Ge atoms are tetrahedrally coordinated by Ni atoms. For MgNiGe, the flat DOS below −8 eV indicates steep bands and thus strong bonding interactions involving the Ge atoms. Similar curve progressions have been reported recently concerning partial DOS of the $E$(s) ($E$ = Ge, Si) orbitals of CaNi$_2$Ge$_2$ and CaCo$_2$Si$_2$ [9, 43]: a flat DOS and thus steep bands are in accordance with the short $E−E$ bonds between the $[T_2E_2]$ layers. On the other hand a sharp partial DOS peak and thus flat bands have been described for the Ge(s) orbitals of BaCo$_2$Ge$_2$ and CaNiGe,[9, 43] which have no Ge−Ge contacts between similar layers.

The narrow band gap of 0.07 eV separates the second and third parts of DOS around −3.1 eV for BaNiGe. In MgNiGe, CaNiGe and SrNiGe this gap is vanished. However, for CaNiGe a pseudogap of 0.1 eV is observed at about −0.8 eV.

Figures 6 and S8-S11 (Supporting Information) show the band structures including fatbands for MgNiGe, CaNiGe SrNiGe and BaNiGe. The difference between the three-dimensional MgNiGe and Sr(Ba)NiGe, with one-dimensional structure motifs, becomes obvious at first sight. In the band structure of MgNiGe band crossing is observed around $E_F$. In contrast, for SrNiGe the bands below and above Fermi level do not mix together, hovewer, they cross the Fermi level $E_F$, what is typically for semimetals. Similarly, such a separation of bands is observed for BaNiGe, however, at the point Γ the bands contact each other. The separation of bands is due to an "avoided crossing" of the bands, which takes place for both SrNiGe and BaNiGe. In order to allow a comparision to the band structure of CaNiGe, a band path comparable to the one used for MgNiGe, SrNiGe and BaNiGe is chosen. Taking into account, that the symmetry reduction from CaNiGe ($P4/mmn$) to Sr(Ba)NiGe ($Pnma$) might lead to avoided crossing of bands, similarities between the band structures become visible. For example, at point $\Gamma$ at −1 eV for CaNiGe bands cross. This might be linked to the avoided crossing of bands at point $\Gamma$ above $E_F$ which is observed for Sr(Ba)NiGe. As the inter-ribbon distances of the $^1_\infty$[NiGe] ribbons are shorter for SrNiGe than for BaNiGe, the separation of the bands is more pronounced. In the band structures of SrNiGe and BaNiGe the avoided crossings next to $E_F$ further are displayed in the local maximum at $E_F$ of the corresponding DOS (Figure 5c-d). No flat bands are observed here. As to be expected for the crystal structure of CaNiGe with two-dimensional structure motifs, no bands cross $E_F$ in the sections parallel to the $c$ axis $\Gamma \rightarrow Z$ and $M \rightarrow A$ are observed. However, one band crosses $E_F$ in the section $R \rightarrow X$, which also corresponds to the same crystallographic direction. The analysis of the fatbands reveals that the orbitals Ca $d_{yz}$, Ni $d_{yz}$ and Ni $d_z^2$ contribute to this band. This underlines the weak interaction of the Ca atoms with the $^2_\infty$[NiGe] layers.

For a more quantitative bonding analysis we performed the crystal orbital Hamilton populations (COHP) which provide a quantitative measure of the strength of the chemical bond. COHP curves for selected interactions are shown in Figures 7 and S12 – S16 (Supporting Information) and the calculated −iCOHP values have been calculated and are given in Table 3. The strongest bonding interactions (i.e. −iCOHP values) were found for the shortest Ni–Ge contacts of the title compounds. For all Ni–Ge contacts a mainly bonding character is found up to Fermi level. The –iCOHP curves clearly reveal that the Ni–Ge bonds are optimized. In contrast, for the Ni–Ni contacts of CaNiGe, SrNiGe and BaNiGe bonding and antibonding interactions are observed below $E_F$. Further, the COHP curves for the *Ae*–Ni contacts are given in Figure S12 – S16 (Supporting Information). They mostly display bonding interactions in the region directly above $E_F$. Further considering the $Ae$–Ni interactions, the calculated –iCOHP values reveal that interactions of cations ($Ae^{2+}$) with [NiGe] framework in MgNiGe is significantly stronger compared to those of CaNiGe and Sr(Ba)NiGe. The –iCOHP values found are to be in good accordance with those of other recently published Ni-rich $Ae$/Ni/Ge compounds.[3, 10-13]

Further insight in the nature of the bonding can be provided by the Electron Localization Function (ELF), sketched in Figure 8. The ELF of BaNiGe has the same general features as observed for SrNiGe, and is therefore neglected. Even though high –iCOHP values are observed for the short Ni–Ge contacts, no disynaptic valence basins are observed here. This is a common feature observed for various compounds of the systems $Ae$/Ni/Ge (see for example [8, 11, 12, 43]). In the MgNiGe two valence basins are observed next to the Ge atoms. The valence basin ① consists of three maxima which appear below $\eta = 0.6$ and are directed towards the neighbouring Mg atoms (Figure 8a). As the four nearest Ni neighbours of Ge atom are arranged in a trigonal pyramid with Ge close to the basal plane, next to Ge a large free space results. Here the basin ① is situated. Valence basin ②, consists of two maxima which appear below $\eta = 0.55$ (Figure 8b) and which are also directed towards neighbouring Mg atoms. For CaNiGe (Figure 8c) one valence basin ③ is observed next to the Ge atoms. It consists of one maxima which is directed towards the nearest Ca atom and which appears below $\eta = 0.78$. For SrNiGe (Figure 8d) one monosynaptic valence basin ④ which appears at $\eta = 0.73$ corresponds to the free electron pair of Ge atom and is oriented in between two neighbouring Sr atoms.

## Conclusion

The size restrictions implied by the alkaline-earth elements Mg, Ca, Sr, and Ba lead to interesting features in the equiatomic $Ae$NiGe. MgNiGe (TiNiSi-type) with three-dimensional $^3_\infty$[NiGe] polyanionic network are formed with the smallest alkaline-earth metal magnesium; CaNiGe crystallizes in well known CeFeSi-type with PbO-like two-dimensional $^2_\infty$[NiGe] layers, whereas the largest Sr and Ba cations cause the formation of structures with one-dimensional $^1_\infty$[NiGe] ribbons. Such one-dimensional polyanions are unique in the crystal chemistry of polar intermetallic compounds and can be considered as the main building block of many transition metals germanides. The substitution of big cation by the smaller one (in Ba–Sr–Ca–Mg row) step by step leads to the compression of the structure, which can be understood as the chemical pressure effects. Therefore, the row of the equiatomic compounds $Ae$NiGe ($Ae$ = Mg, Ca, Sr, Ba) is an excellent example of the influence of the $Ae$ cation size on the dimensionality of the polyanions.

It is tempting to associate the structure to a Zintl electronic scheme as $Ae^{2+}Ni^{2+}Ge^{4-}$. However, the close electronegativities of Ni and Ge, as well as existence of significant Ni–Ge interactions dispute this simple assignment. It is more appropriate to depict the [NiGe] infinite 1D-ribbon, 2D-layer and 3D-network as $[NiGe]^{2-}$ polyanions. As pointed out, the size of cations (going from Mg to Ba) plays crucial role on the structure formation in case of alkaline-earth nickel germanides. Nevertheless, other equiatomic alkaline-earth transition metal germanides SrMnGe and BaMnGe [44] with large cations, as well as MgCuGe [45] and recently described MgCoGe [9] with small cations adopt the same tetragonal CeFeSi type with 2-D polyanionic layers, comparable to the title compound CaNiGe. Equiatomic rare-earth metal nickel germanides crystallize with orthorombic TiNiSi (or CeCu$_2$) structure types with 3-D polyanionic networks [27,46], whereas EuNiGe (isoelectronic to $Ae$NiGe) adopts monoclinic $CoSb_2$ type with 2-D polyanionic layers [47, 48]. All these facts of structural diversity suggest, firstly, that the formation of the structure types depends − besides of atomic size − also on the valence electron concentration. Secondly − the possible polymorphism in described compounds as was described previously for other equiatomic germanides such as CaAuGe [49], LuNiGe [50] and *Re*PdGe [51].

Furthermore, the existence of isolated $^1_\infty$[NiGe] ribbons in Sr(Ba)NiGe as well as their prevalence in other intermetallic compounds opens the possibilities of using such building blocks as precursors for the new composite materials.

## Supporting Information

File containing detailed crystallographic data of crystals MgNiGe, SrNiGe, and BaNiGe in CIF format; figures of X-ray powder patterns of MgNiGe, SrNiGe, and BaNiGe (Figures S1−S3); the structures of other transition metal germanides containing the $^1_\infty$[*T*Ge] ribbons (Figure S4); the structures of $PbCl_2$, PbFCl, and SnFCl (Figure S5); temperature dependence of the magnetic susceptibility of BaNiGe and SrNiGe (Figure S6); partial DOS curves for the alkaline-earth metal s, p and d orbitals of *Ae*NiGe (*Ae* = Mg, Ca, Sr, Ba) (Figure S7); band structure including fatbands of *Ae*NiGe (*Ae* = Mg, Ca, Sr, Ba) (Figures S8−S11); COHP of the Ni−Ni, *Ae*−Ni and *Ae*−Ge interaction for *Ae*NiGe (*Ae* = Mg, Ca, Sr, Ba) (Figures S12−S16). This material is available free of charge via the Internet at http://pubs.acs.org.

## Acknowledgments

This research was supported by the Deutsche Forschungsgemeinschaft within the priority program SPP-1458 (project HL 62/1-1).

**Table 1.** Crystal data and structure refinement for MgNiGe, SrNiGe, and BaNiGe (space group *Pnma*, Z=4)

| empirical formula | MgNiGe | SrNiGe | BaNiGe |
|---|---|---|---|
| $M$w, g mol$^{-1}$ | 155.61 | 218.92 | 268.64 |
| unit cell parameters | | | |
| (powder data):    *a*, Å | 6.4742(2) | 5.727(1) | 5.969(4) |
| *b*, Å | 4.0716(1) | 4.174(1) | 4.195(1) |
| *c*, Å | 6.9426(2) | 11.400(3) | 11.993(5) |
| *V*, Å$^3$ | 183.01(1) | 272.5(1) | 300.3(2) |
| $\mu$ (Mo K$\alpha$), mm$^{-1}$ | 26.5 | 36.9 | 28.8 |
| $\rho_{calc}$, g cm$^{-3}$ | 5.647 | 5.336 | 5.942 |
| diffractometer | Oxford-Xcalibur3 | Stoe IPDS-2T | Stoe IPDS-2T |
| crystal size (mm$^3$) | 0.04×0.07×0.09 | 0.01×0.02×0.08 | 0.01×0.02×0.06 |
| $F(000)$ | 288 | 392 | 464 |
| $\theta$ range for data collection (deg) | 3 – 30 | 5 – 27 | 2 – 32 |
| range in *hkl* | $-9 \leq h \leq 8$, $-5 \leq k \leq$ 4, $\pm 9$ | $\pm 7$, $\pm 5$, $\pm 14$ | $\pm 7$, $\pm 5$, $\pm 15$ |
| reflections collected | 1556 | 4201 | 4590 |
| independent reflections | 305 ($R_{int} = 0.016$) | 358 ($R_{int} = 0.110$) | 393 ($R_{int} = 0.061$) |
| reflections with $I \geq 2\sigma(I)$ | 294 ($R_{sigma} = 0.010$) | 298 ($R_{sigma} = 0.040$) | 354 ($R_{sigma} = 0.021$) |
| data / parameters | 305 / 20 | 358 / 20 | 393 / 20 |
| GOF on $F^2$ | 1.298 | 1.057 | 1.077 |
| $R_1$, $wR_2$ [$I \geq 2\sigma(I)$] | 0.015, 0.033 | 0.036, 0.078 | 0.022, 0.048 |
| $R_1$, $wR_2$ (all data) | 0.015, 0.033 | 0.048, 0.082 | 0.027, 0.049 |
| extinction coefficient | 0.026(2) | 0.003(1) | 0.0024(5) |
| largest diff. peak and hole (e Å$^{-3}$) | 0.58 and $-0.65$ | 1.04 / $-1.08$ | 1.97 / $-0.92$ |

[a] $R_1 = \Sigma ||F_c| - |F_c|| / \Sigma |F_o|$; $wR_2 = [\Sigma[w(F_o^2 - F_c^2)2] / \Sigma[w(F_o^2)^2]]^{1/2}$; [b] $w = 1/[\sigma^2(F_o^2) + (aP)2 + bP]$, where $P = (\text{Max}(F_o^2, 0) + 2F_c^2)/3$.

**Table 2.** Atomic coordinates and equivalent isotropic displacement parameters ($\text{Å}^2 \times 10^3$) for SrNiGe and BaNiGe. $U_{eq}$ is defined as one third of the trace of the orthogonalized $U_{ij}$ tensor.

| Atom | Wyckoff position | $x$ | $y$ | $z$ | $U_{eq}$ |
|------|------------------|-----|-----|-----|----------|
| **MgNiGe** | | | | | |
| Mg | 4$c$ | 0.0019(2) | 1/4 | 0.6876(2) | 12(0) |
| Ni | 4$c$ | 0.34605(6) | 1/4 | 0.44589(5) | 10(0) |
| Ge | 4$c$ | 0.21043(5) | 1/4 | 0.11862(4) | 9(0) |
| | | | | | |
| **SrNiGe** | | | | | |
| Sr | 4$c$ | 0.1194(2) | 1/4 | 0.35634(9) | 26(1) |
| Ni | 4$c$ | 0.1006(2) | 1/4 | 0.0441(1) | 23(1) |
| Ge | 4$c$ | 0.3646(2) | 1/4 | 0.6288(1) | 25(1) |
| | | | | | |
| **BaNiGe** | | | | | |
| Ba | 4$c$ | 0.11787(7) | 1/4 | 0.35200(3) | 21(0) |
| Ni | 4$c$ | 0.0960(2) | 1/4 | 0.04111(7) | 19(0) |
| Ge | 4$c$ | 0.3690(1) | 1/4 | 0.62128(6) | 20(1) |

**Table 3.** Interatomic distances (Å), calculated with the lattice parameters taken from X-ray powder data and corresponding integrated crystal orbital Hamilton populations (–iCOHPs) values at $E_F$ of MgNiGe, SrNiGe and BaNiGe. All distances within the first coordination spheres are listed. All -iCOHP values are in eV per bond.

| | | distance (Å) | –iCOHP (eV/bond) | | | distance (Å) | –iCOHP (eV/bond) |
|---|---|---|---|---|---|---|---|
| **MgNiGe** | | | | | | | |
| Ni | Ge | 2.391(1) (2×) | 2.13 | Mg | Ni | 2.737(2) (1×) | 0.67 |
| | Ge | 2.401(1) (1×) | 2.00 | | Ni | 2.789(2) (1×) | 0.84 |
| | Ge | 2.436(1) (1×) | 2.22 | | Ge | 2.800(1) (2×) | 0.72 |
| | | | | | Ge | 2.801(1) (2×) | 0.68 |
| Ge | Ni | 2.391(1) (2×) | 2.13 | | Ge | 2.843(2) (1×) | 0.70 |
| | Ni | 2.401(1) (1×) | 2.00 | | Ni | 2.886(1) (2×) | 0.62 |
| | Ni | 2.436(1) (1×) | 2.22 | | Ge | 3.283(1) (2×) | 0.32 |
| | | | | | | | |
| **SrNiGe** | | | | | | | |
| Ni | Ge | 2.308(2) (2×) | 2.95 | Sr | Ni | 2.980(2) (1×) | 0.46 |
| | Ge | 2.390(2) (1×) | 2.11 | | Ni | 3.180(2) (1×) | 0.47 |
| | Ni | 2.588(2) (2×) | 1.10 | | Ge | 3.331(2) (2×) | 0.55 |
| | | | | | Ni | 3.393(2) (2×) | 0.28 |
| Ge | Ni | 2.308(2) (2×) | 2.95 | | Ge | 3.409(2) (1×) | 0.30 |
| | Ni | 2.390(2) (1×) | 2.11 | | Ge | 3.474(2) (2×) | 0.36 |
| | | | | | Ni | 3.561(2) (1×) | 0.22 |
| | | | | | Ge | 3.622(2) (2×) | 0.24 |
| | | | | | | | |
| **BaNiGe** | | | | | | | |
| Ni | Ge | 2.317(1) (2×) | 2.97 | Ba | Ni | 3.129(2) (1×) | 0.49 |
| | Ge | 2.372(2) (1×) | 2.26 | | Ni | 3.369(3) (1×) | 0.45 |
| | Ni | 2.585(2) (2×) | 1.13 | | Ge | 3.473(2) (2×) | 0.56 |
| | | | | | Ni | 3.530(2) (2×) | 0.29 |
| Ge | Ni | 2.317(1) (2×) | 2.97 | | Ge | 3.560(2) (1×) | 0.29 |
| | Ni | 2.372(2) (1×) | 2.26 | | Ge | 3.598(2) (2×) | 0.37 |
| | | | | | Ge | 3.726(2) (2×) | 0.26 |
| | | | | | Ni | 3.732(2) (1×) | 0.22 |

**Table 4.** Inter- and intra-ribbon Ni-Ge and Ni-Ni distances as well as the angle of the pleated sheet structure of MgNiGe, CaNiGe, SrNiGe, BaNiGe, $Ca_4Ni_4Ge_3$, $SrNi_2Ge$, $BaNi_2Ge$ and $SrNi_3Ge_2$.

| | $d$(Ni-Ge) intra-ribbon /Å | $d$(Ni-Ge) inter-ribbon /Å | $d$(Ni-Ni) intra-ribbon /Å | $d$(Ni-Ni) inter-ribbon /Å | $\alpha$[a] /° |
|---|---|---|---|---|---|
| **MgNiGe** | 2.391(1) 2.301(1) | 2.436(1) | 2.947(1) | - | 122.80(2) |
| **CaNiGe** | 2.433 | 2.433 | 2.965(1) | 2.965 | 126.20 |
| **SrNiGe** | 2.308(1) 2.390(2) | - | 2.588(2) | - | 159.30(7) |
| **BaNiGe** | 2.317(1) 2.372(2) | - | 2.586(2) | - | 159.60(5) |
| **$Ca_4Ni_4Ge_3$** | 2.407 2.416 | 2.451 [b] | 2.526 | 2.487 | 151.90 |
| **$SrNi_2Ge$** | 2.409 | 2.409 [c] | 2.485 | 2.484 [c] | 151.69 |
| **$BaNi_2Ge$** | 2.422 2.428 | 2.422 [c] 2.428 [c] | 2.533 | 2.483 [c] | 150.03 |
| **$SrNi_3Ge_2$** | 2.395 2.362 | 2.362 2.706 [d] | 3.121 | 3.121 2.557 [d] | 120.00 |

a) Dihedral angle α of the pleated sheet structure (See Sheme 1)

b) Distance ribbon – (Ni-Ge-Ni) bridge, see Figure 2d

c) Distance ribbon – Ni chain, see Figure 2e, f

d) Distance ribbon – Ni atom, see Figure 2g

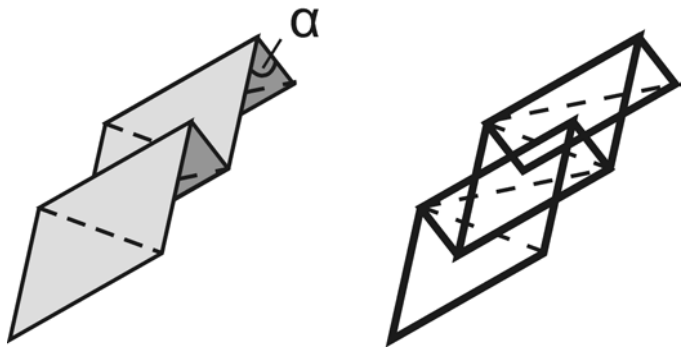

**Sheme 1.** Pleated sheet $^1_\infty$[NiGe]. The Ni-Ni bonds are shown as dashed lines, the dihedral angle α is given.

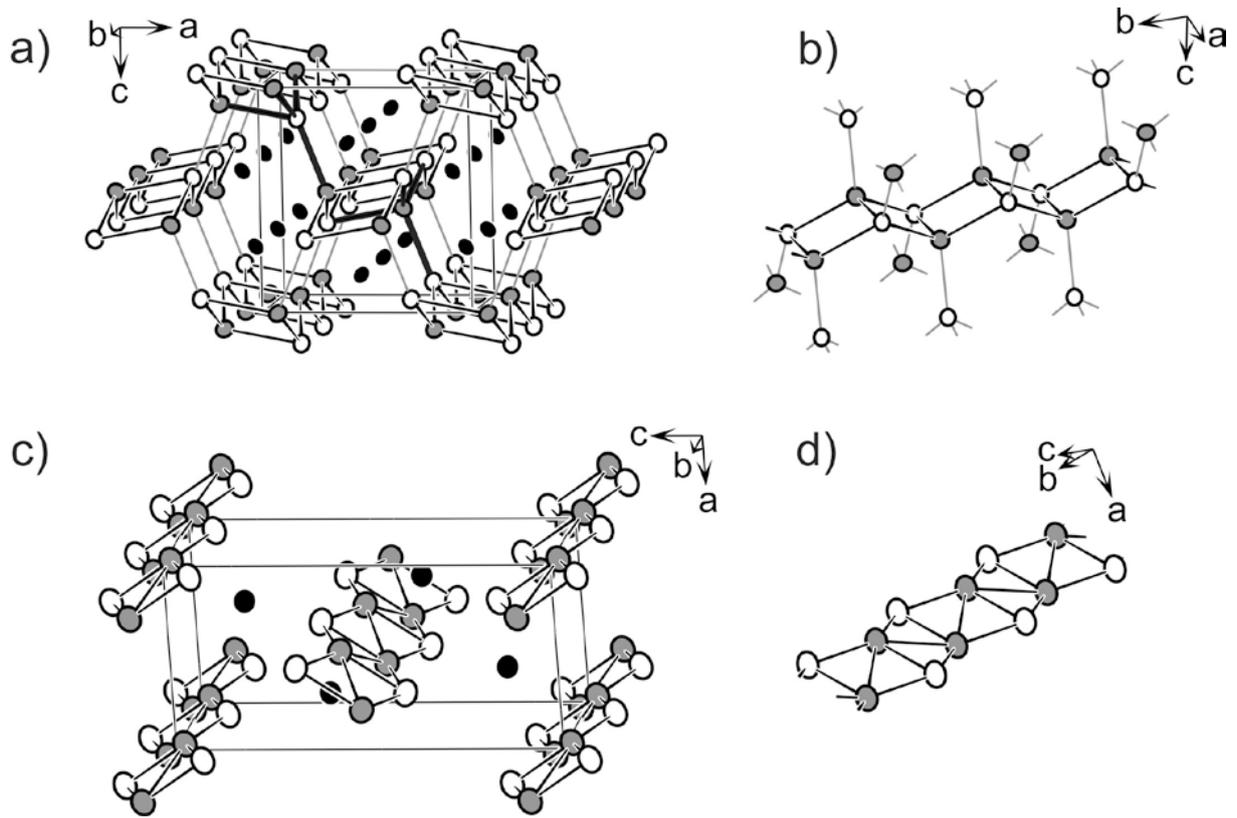

**Figure 1.** View of the a) MgNiGe and c) BaNiGe structures approximately along the *b* axis. The three-dimensional $^3_\infty$[NiGe] network of MgNiGe as well as one-dimensional $^1_\infty$[NiGe] ribbons of BaNiGe are emphasized. For MgNiGe the coordination environment of Ni and Ge is emphazised. A closer look on the $^1_\infty$[NiGe] substructures is given in b) for MgNiGe and d) for BaNiGe. The Mg (Ba) atoms are drawn as black spheres, Ni and Ge atoms as grey and white ones, respectively. The displacement ellipsoids are drawn with a 90% probability level.

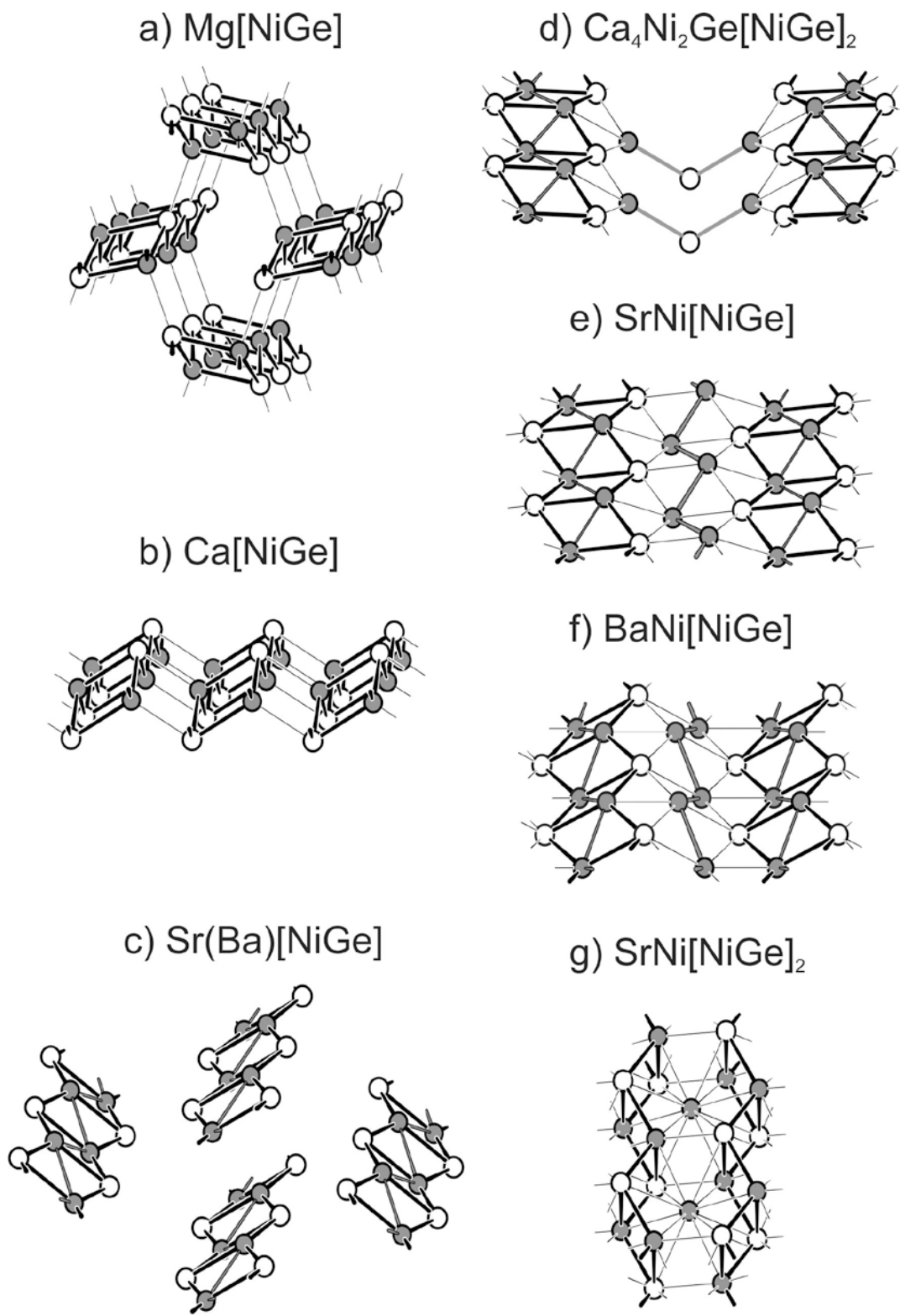

**Figure 2.** Ni-Ge substructures of a) MgNiGe, b) CaNiGe, c) Sr(Ba)NiGe, d) Ca$_4$Ni$_4$Ge$_3$, e) SrNi$_2$Ge, f) BaNi$_2$Ge, g) SrNi$_3$Ge$_2$ containing similar $^1_\infty$[NiGe] ribbons (emphasized). The Ni and Ge atoms are drawn in grey and white, respectively.

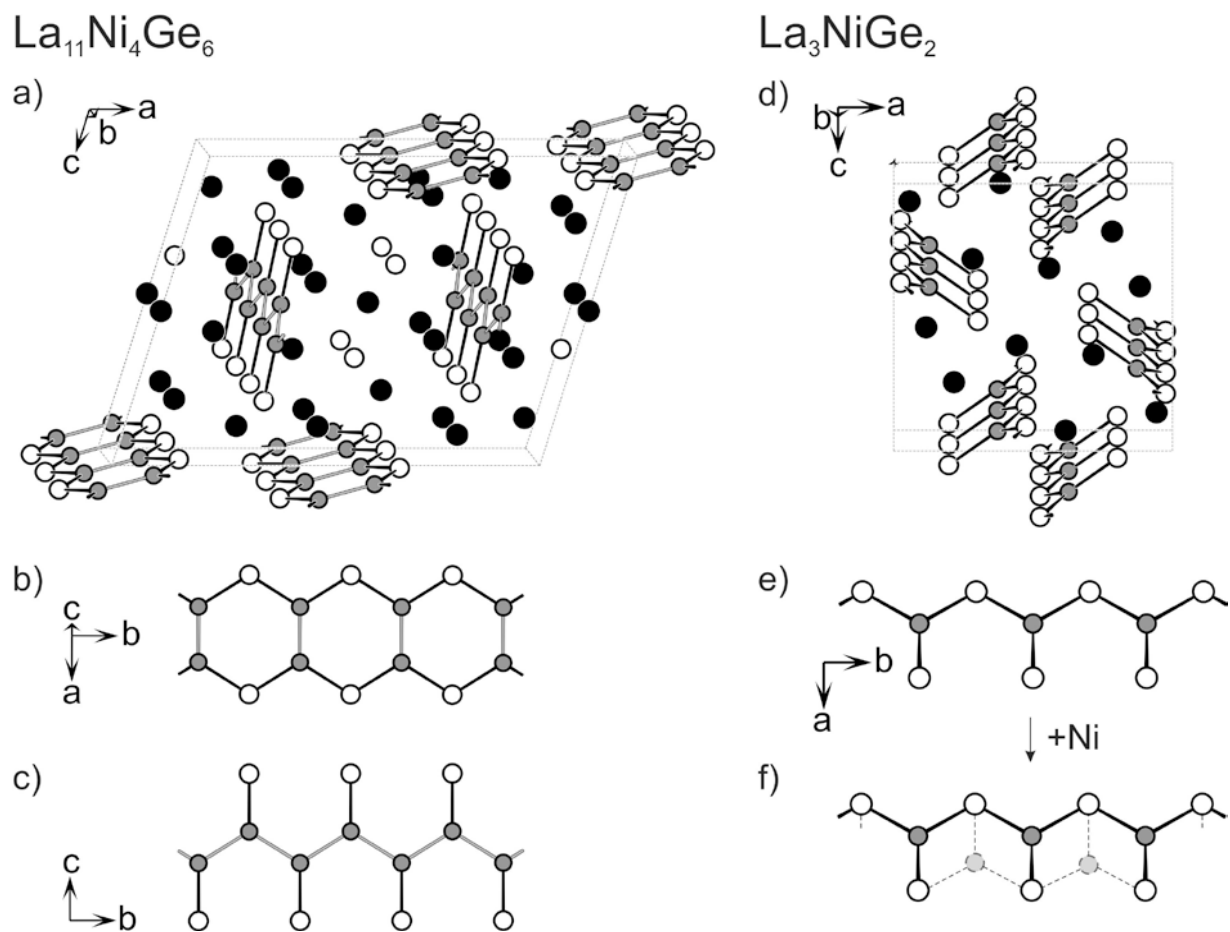

**Figure 3.** Crystal structure of a) $La_{11}Ni_4Ge_6$ with one-dimensional polymorphic $^1_\infty$[NiGe] ribbons (b and c). Crystal structure of d) $La_3NiGe_2$ containing one-dimensional $^1_\infty$[NiGe$_2$] chain (e), which is a defective variant of the $^1_\infty$[NiGe] ribbon (f) of Sr(Ba)NiGe. The Ni, Ge, and La atoms are drawn in grey, white, and black, respectively.

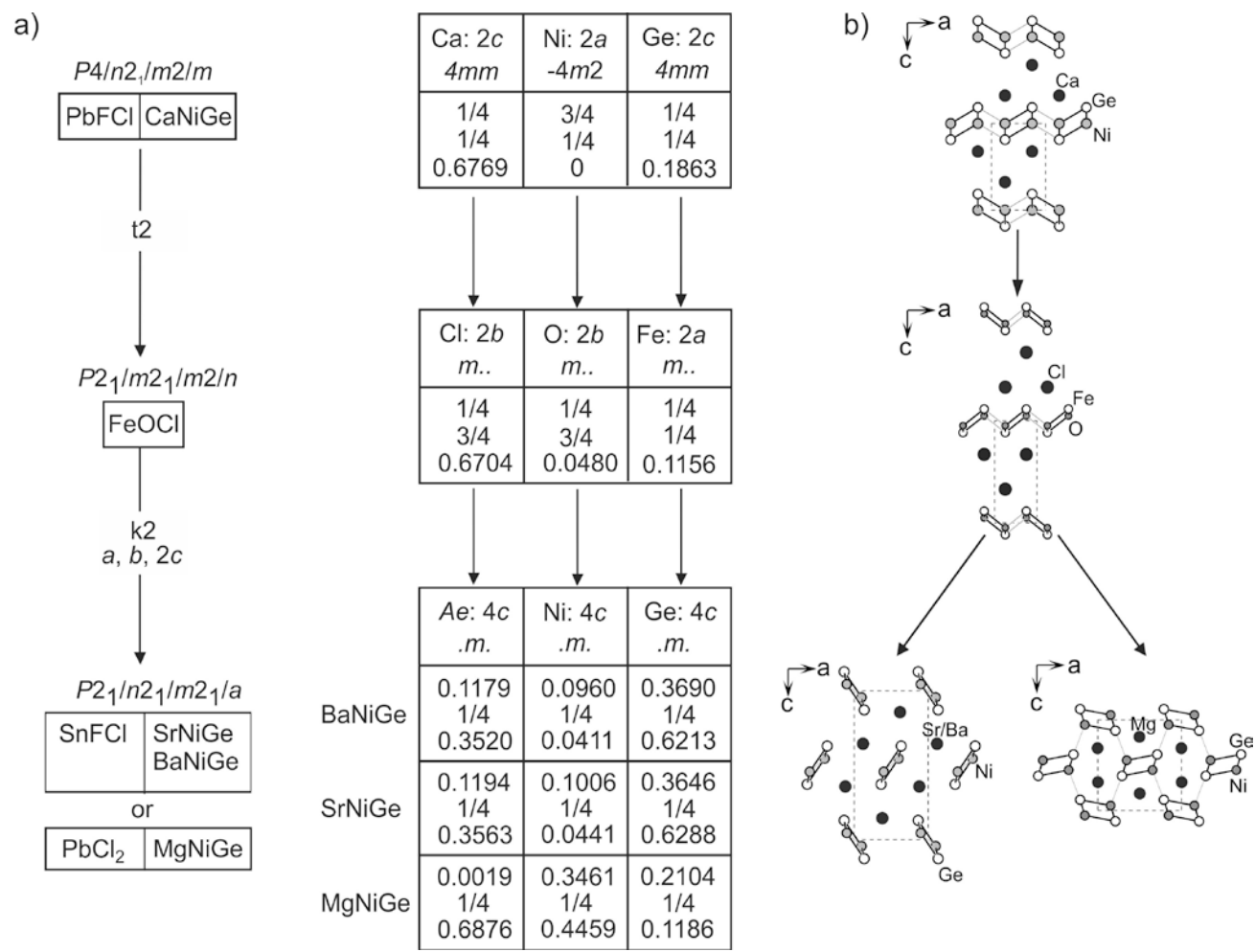

**Figure 4.** a) Group-subgroup relation for the structures of CaNiGe (PbFCl), FeOCl, Sr(Ba)NiGe (SnFCl), and MgNiGe (PbCl$_2$). The indices for the *translationengleiche* (t), *klassengleiche* (k), isomorphic (i) transitions and the unit cell transformations are given together with the evolution of the atomic parameters; b) representations of corresponding structures. The crystal structures of PbFCl, SnFCl and PbCl$_2$ are given in Supporting Information, Figure S5.

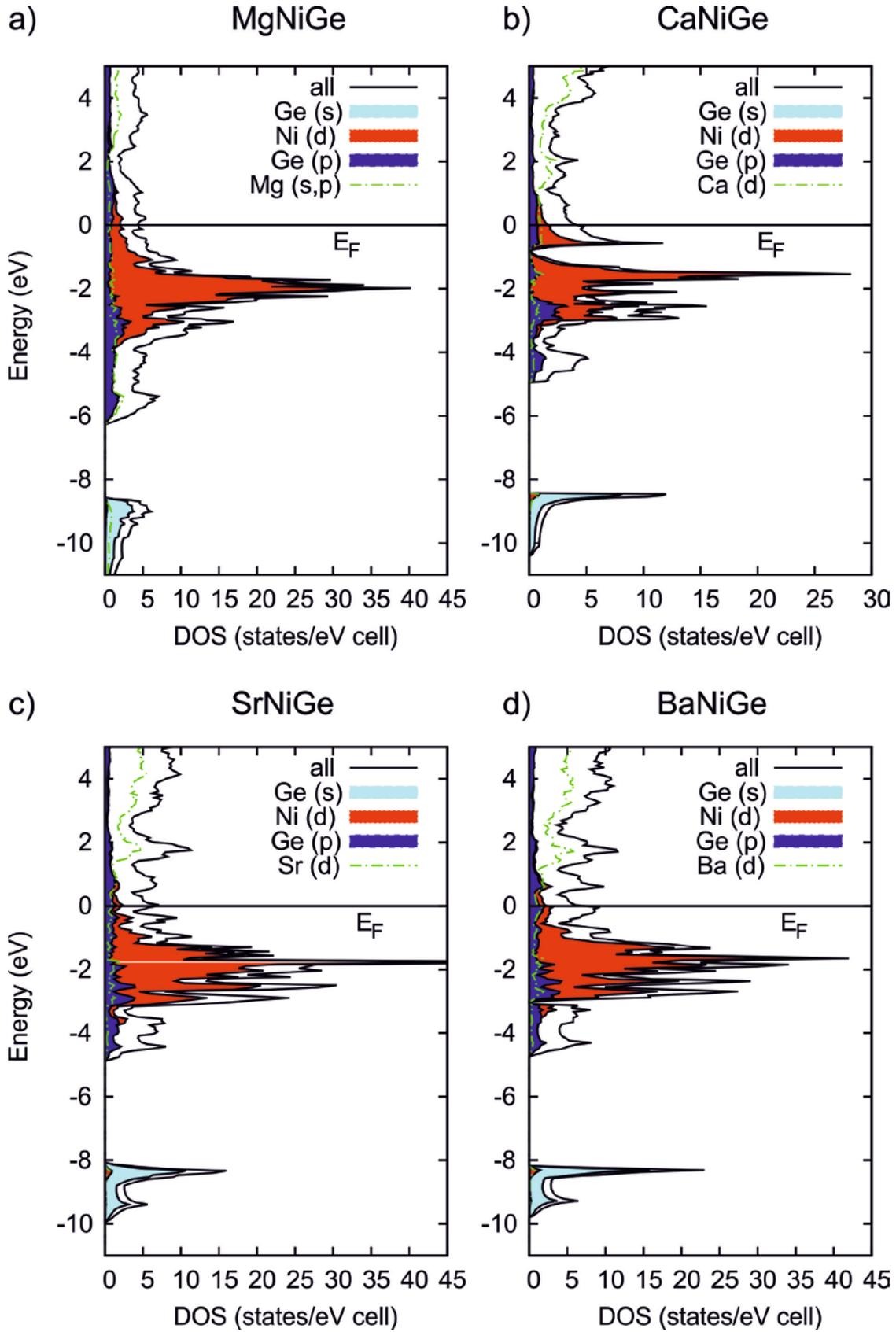

**Figure 5.** Total and partial DOS calculated for a) MgNiGe, b) CaNiGe, c) SrNiGe and d) BaNiGe. The energy zero is taken at the Fermi level.

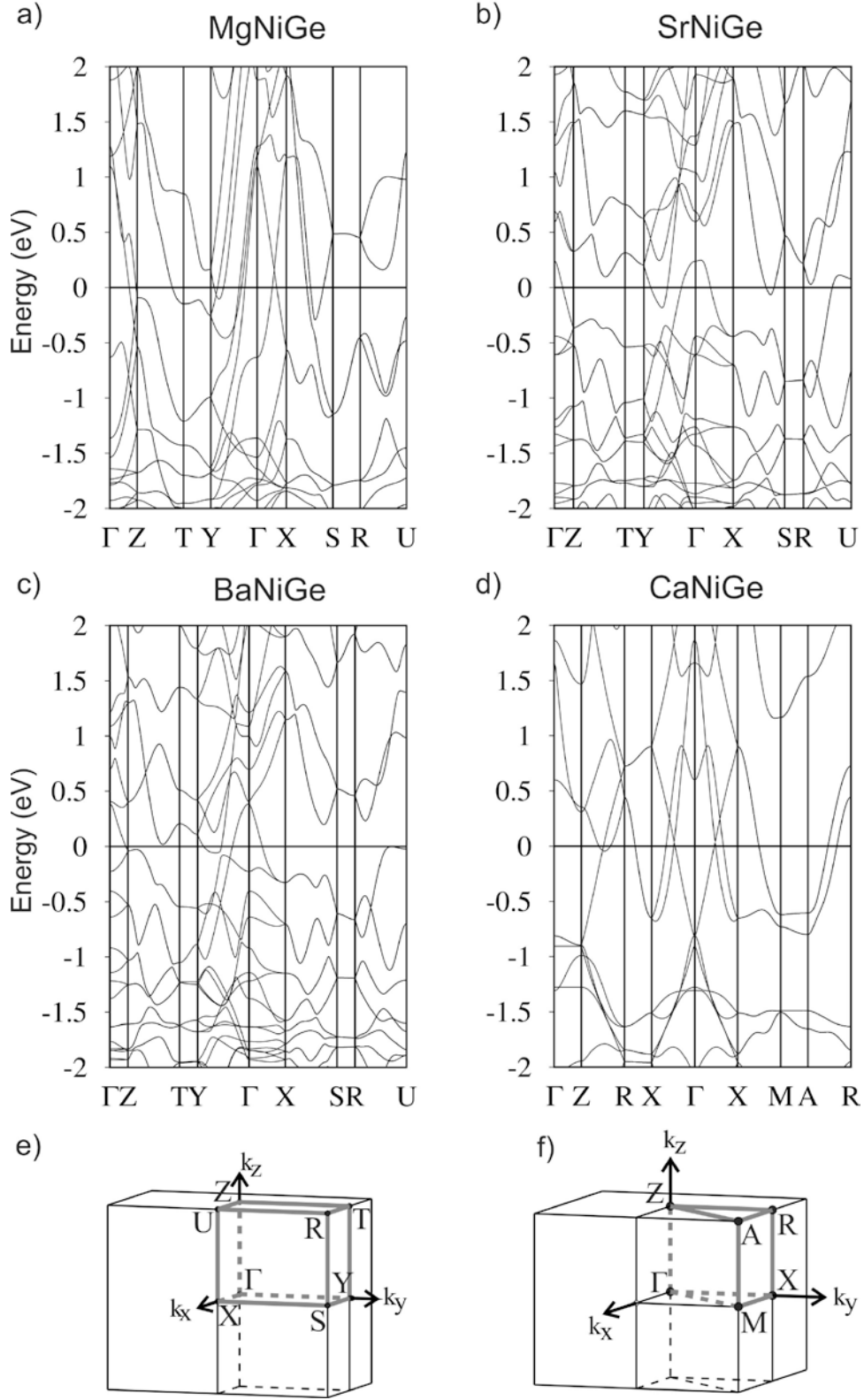

**Figure 6.** Band structure for a) MgNiGe, b) SrNiGe, c) BaNiGe and d) CaNiGe in the range from −2 eV to 2 eV. The symmetry points in *k* space are given according to the Brillouin zone with respect to the reciprocal conventional vectors e) for space group *Pnma* (MgNiGe, SrNiGe and BaNiGe), f) for space group *P4/nmm* (CaNiGe).

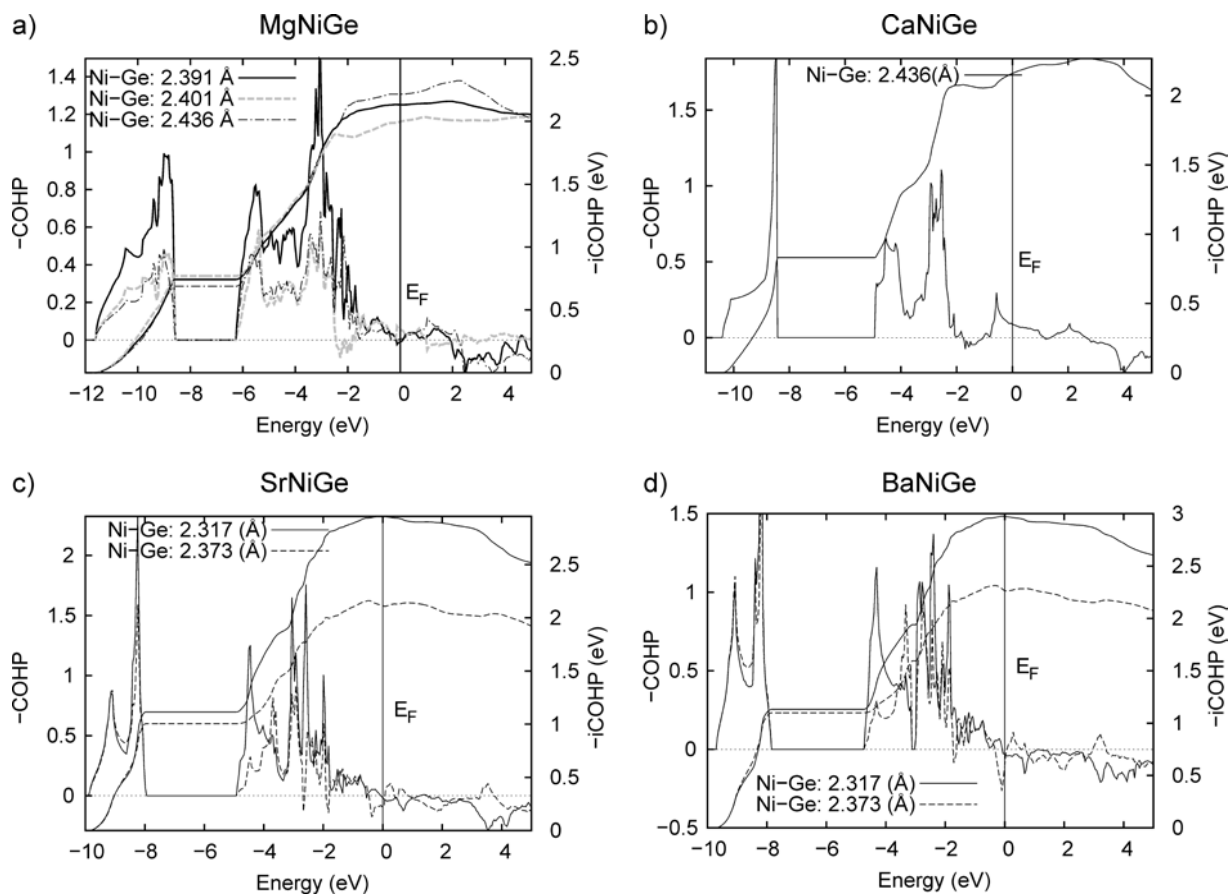

**Figure 7.** Crystal Orbital Hamilton Populations (COHP) and integrated crystal orbital Hamilton populations (−iCOHP) curves for corresponding Ni-Ge and Ni-Ni bonds within the [NiGe] substructure of MgNiGe (a), CaNiGe (b), SrNiGe (c) and BaNiGe (d).

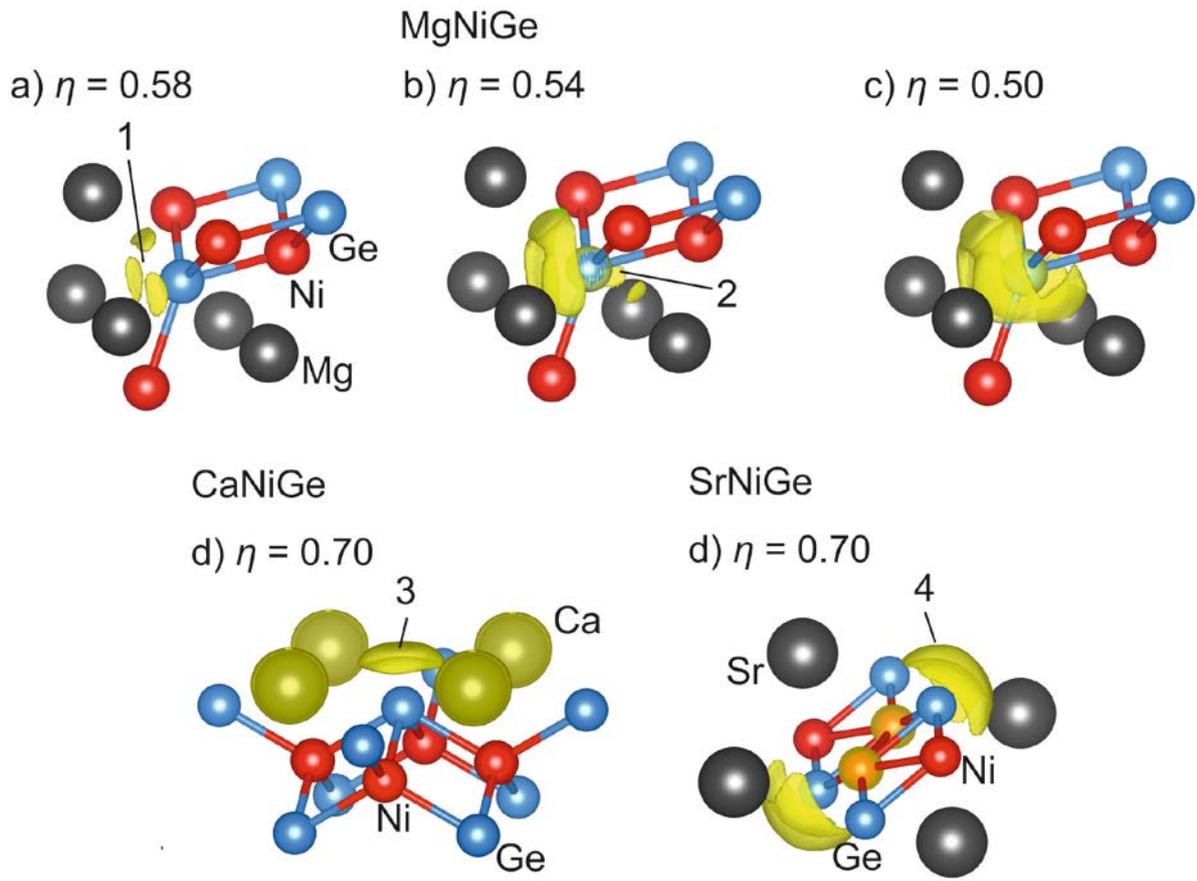

**Figure 8.** Three dimensional ELF plot with isosurface at $\eta(r) = 0.58$ (a), 0.54 (b), 0.50 (c) for MgNiGe; d) isosurface at $\eta(r) = 0.70$ for CaNiG; e) isosurface at $\eta(r) = 0.70$ for SrNiGe.

# Supporting Information

# From one to three dimensions – Corrugated $^1_\infty$[NiGe] ribbons as building block in alkaline-earth metal *Ae/*Ni/Ge phases. Crystal structure and chemical bonding in *Ae*NiGe (*Ae* = Mg, Sr, Ba)

**Viktor Hlukhyy\*, Lisa Siggelkow, Thomas F. Fässler**

Departement of Chemistry, Technische Universität München, Lichtenbergstr. 4, 85747 Garching , Germany

**Address for correspondence:**

\* Dr. Viktor Hlukhyy, Department of Chemistry, Technische Universität München, Lichtenbergstr. 4, D-85747 Garching Germany, E-mail: viktor.hlukhyy@lrz.tu-muenchen.de

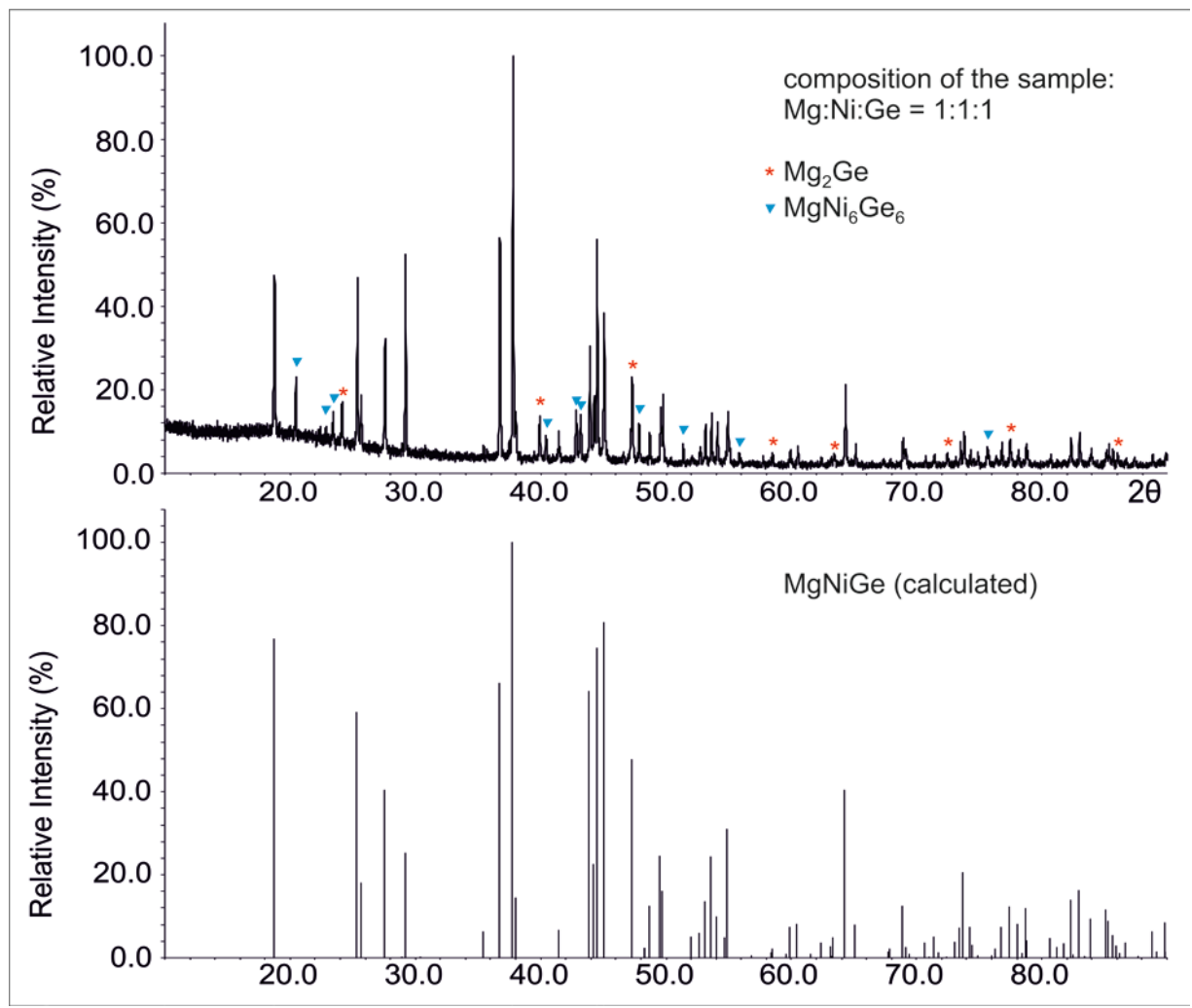

**Figure S1.** Experimental XRD powder pattern from the MgNiGe sample (top) and simulated powder xrd diagram of MgNiGe (bottom). The experimental XRD powder pattern was recorded in transmission geometry. Reflections of Mg$_2$Ge and MgNi$_6$Ge$_6$ are labeled with a red stars and blue triangles, respectively.

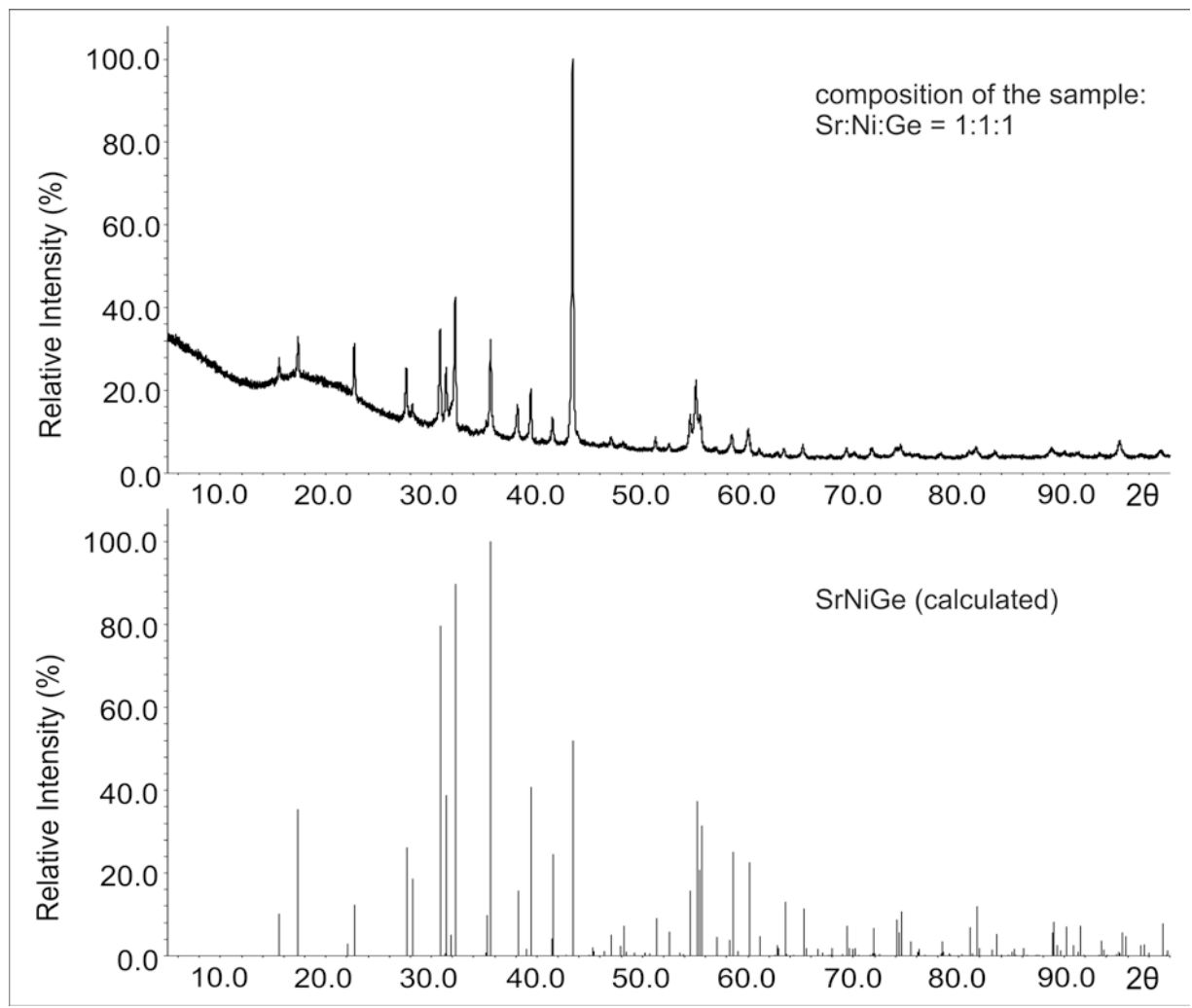

**Figure S2.** Experimental XRD powder pattern from the SrNiGe sample (top) and simulated powder XRD pattern of SrNiGe (bottom). The experimental XRD powder pattern was recorded in transmission geometry.

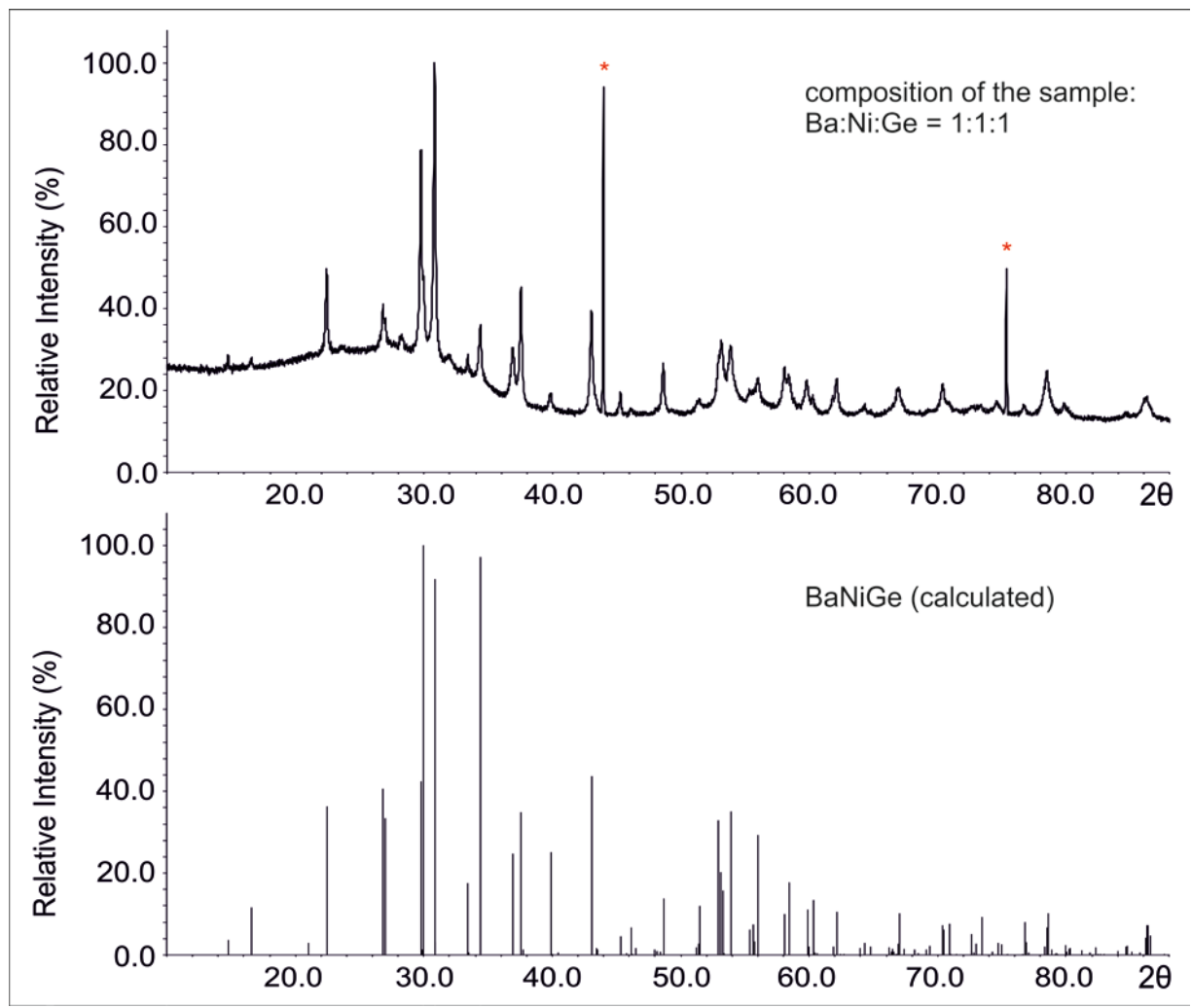

**Figure S3.** Experimental XRD powder pattern from the BaNiGe sample (top) and simulated powder XRD pattern of BaNiGe (bottom). The experimental XRD powder pattern was recorded in transmission geometry. Reflections of diamond are labeled with a red stars.

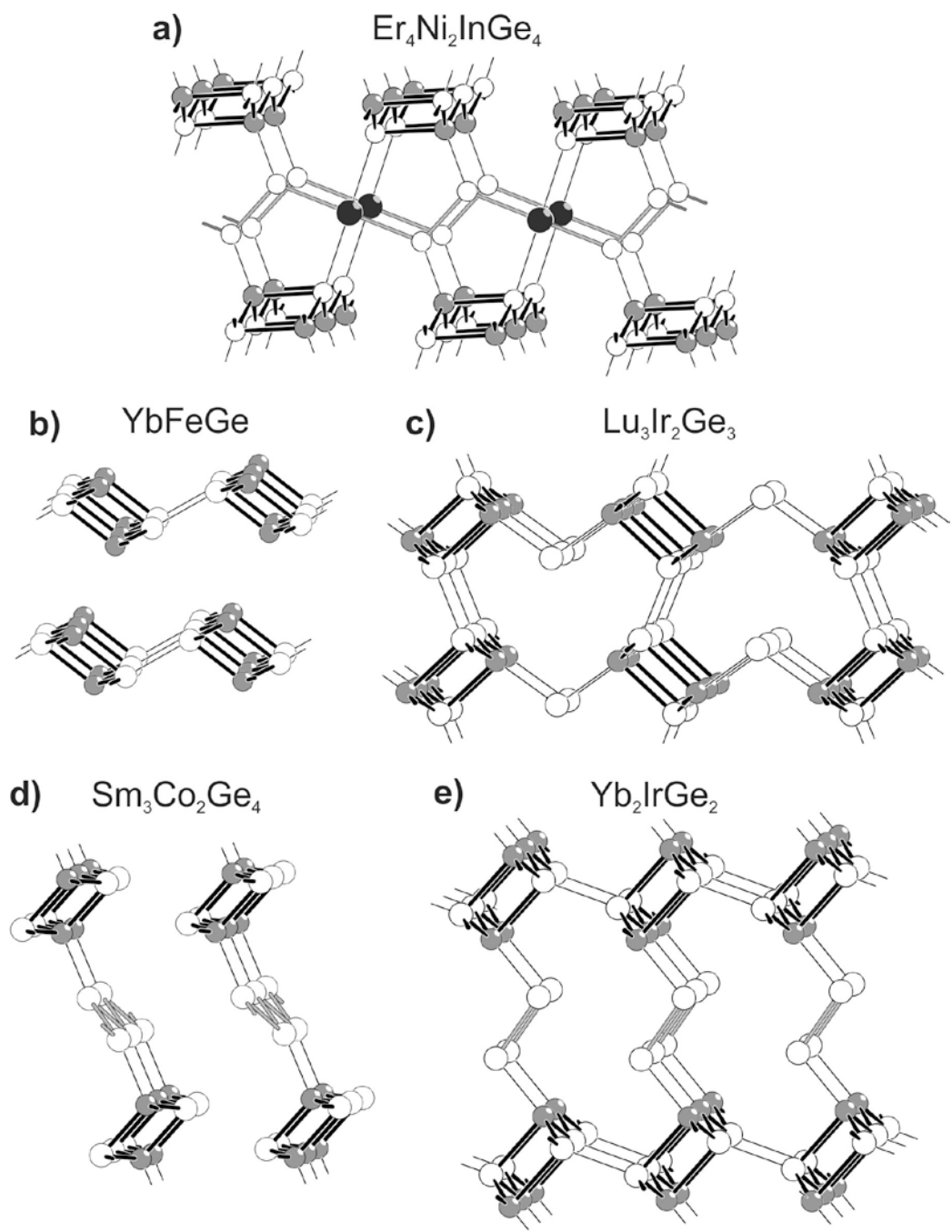

**Figure S4.** *T*-Ge substructures of Er$_4$Ni$_2$InGe$_4$ (a), YbFeGe (b), Lu$_3$Ir$_2$Ge$_3$ (c), Sm$_3$Co$_2$Ge$_4$ (d), Yb$_2$IrGe$_2$ (e), containing [*T*Ge] ribbons. The *T*, Ge, and In atoms are drawn in grey, white, and black, respectively.

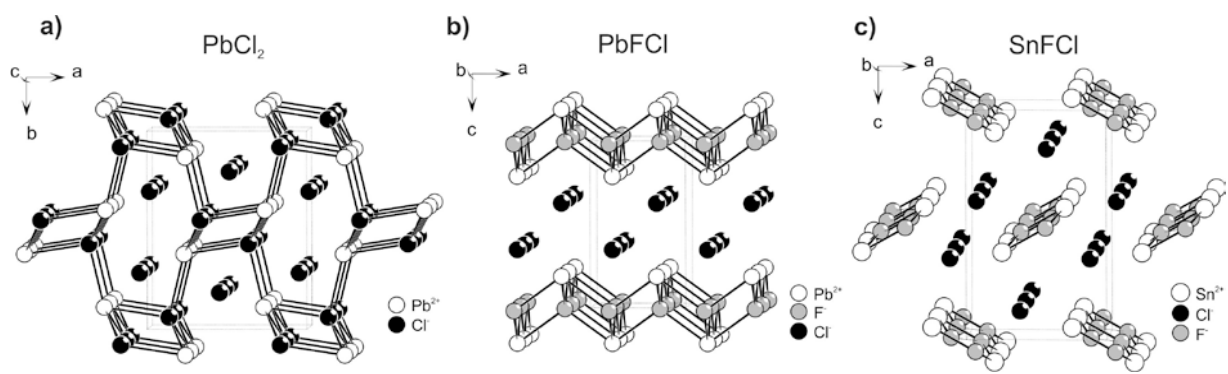

**Figure S5.** The structures of a) PbCl$_2$, b) PbFCl, and c) SnFCl. The Pb-Cl, Pb-F, and Sn-F contacts in PbCl$_2$, PbFCl, and SnFCl are drawn in order to compare with Ni-Ge polyanions of Sr(Ba)NiGe, CaNiGe, and MgNiGe, respectively (see Figures 1-2 ).

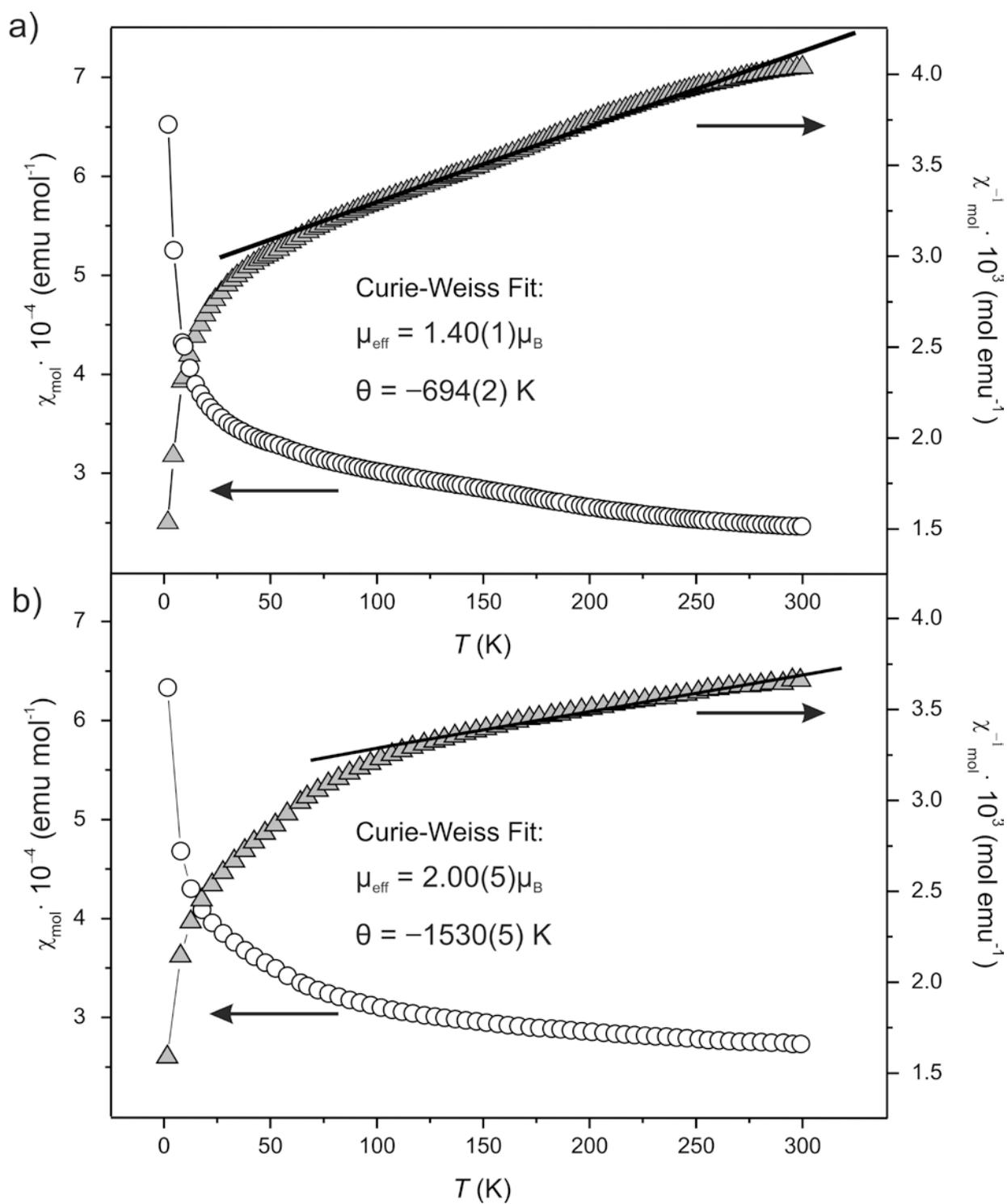

**Figure S6.** Temperature dependence of the magnetic susceptibility of BaNiGe (a) and SrNiGe (b), measured on polycrystalline sample using an applied magnetic field $H$ = 5 kOe and 1 kOe, respectively.

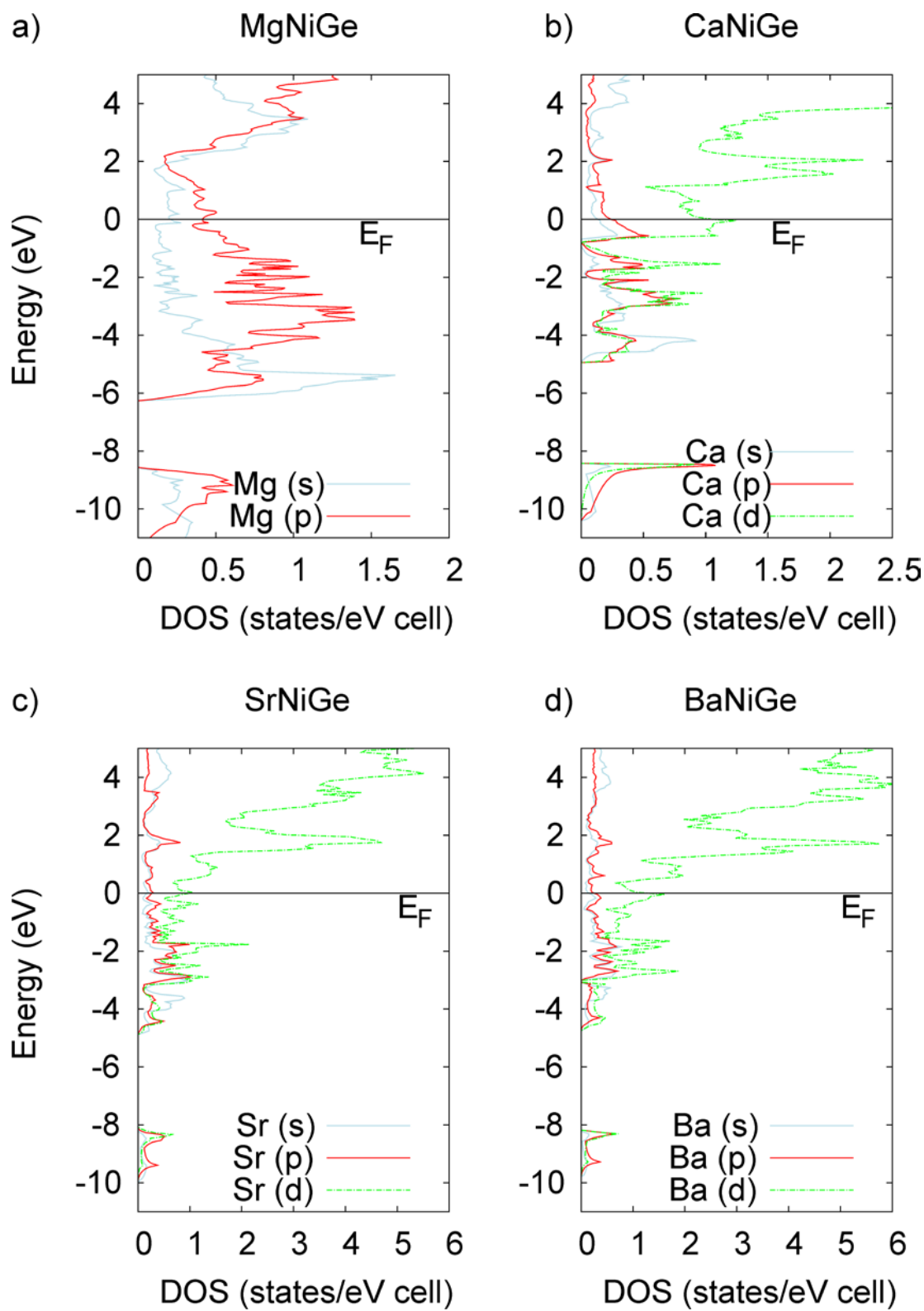

**Figure S7.** Partial DOS curves for the alkaline-earth metal s, p and d orbitals of a) MgNiGe, b) CaNiGe, c) SrNiGe, d) BaNiGe

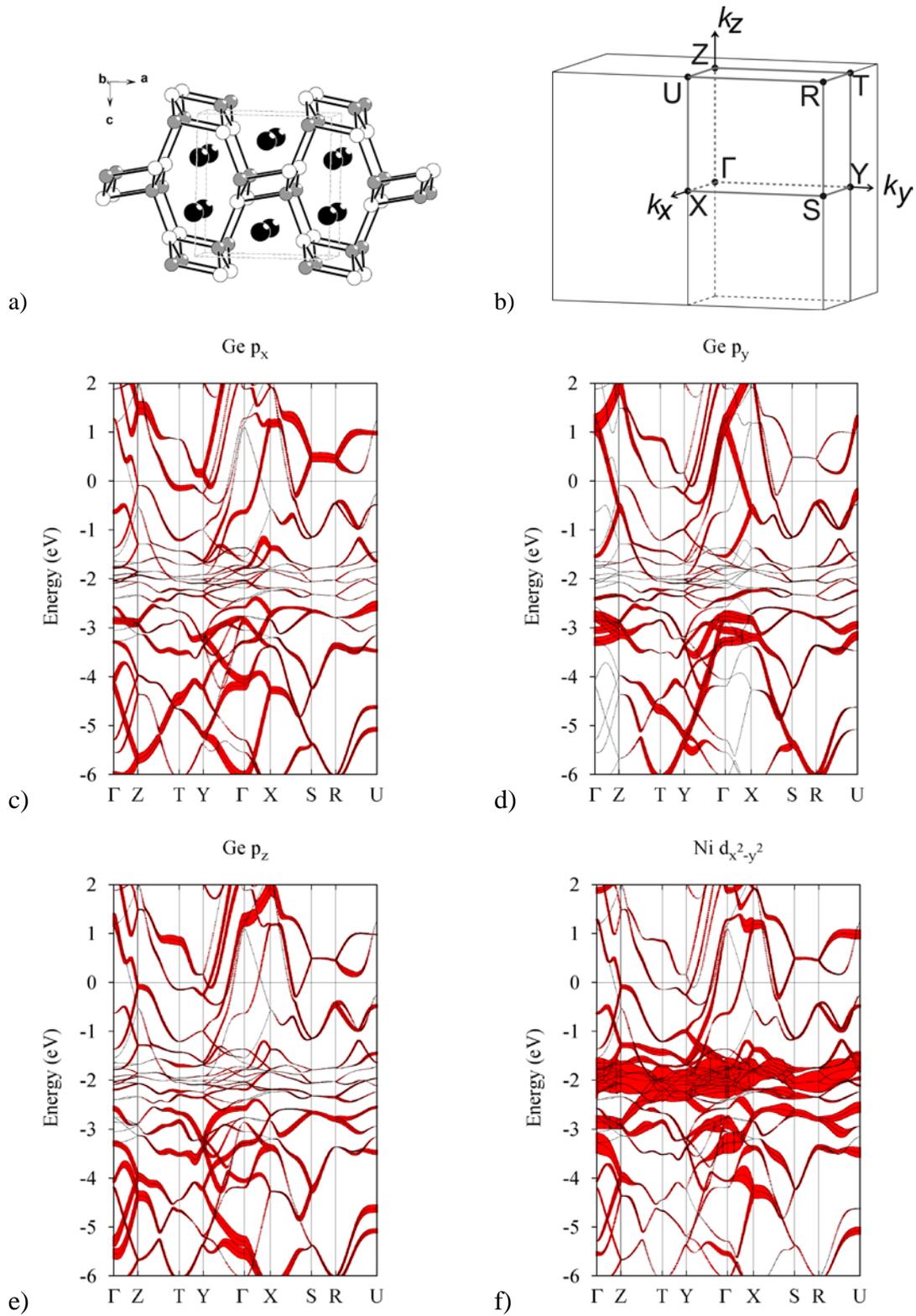

**Figure S8.** Band structure including fatbands of MgNiGe.

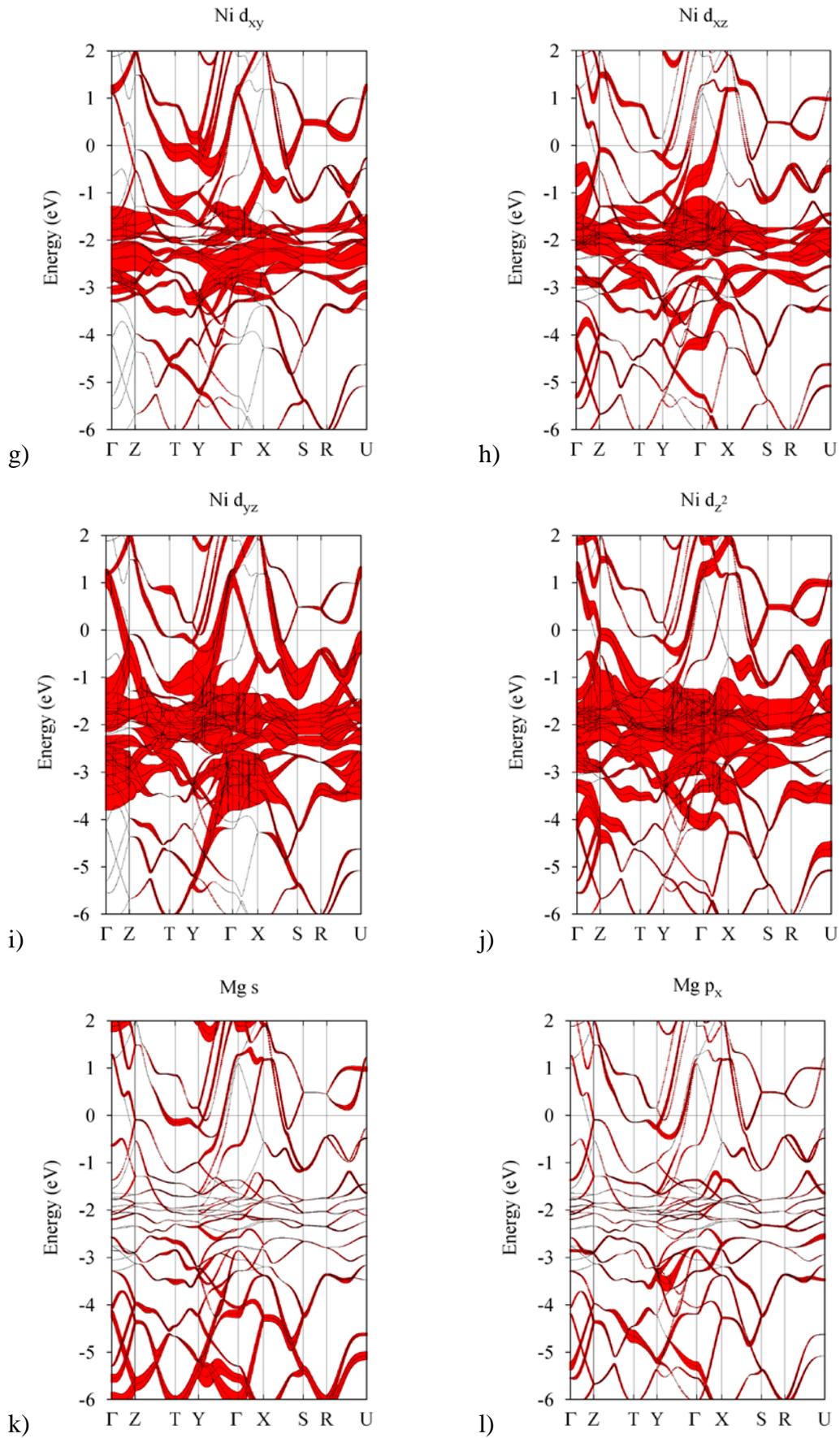

**Figure S8 (continued).** Band structure including fatbands of MgNiGe.

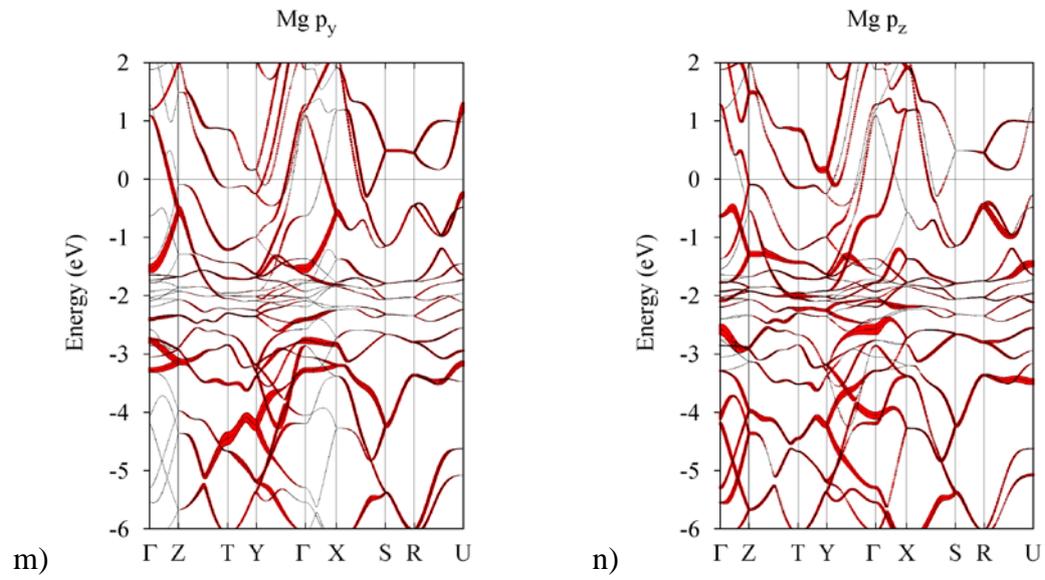

m)          n)

**Figure S8 (continued).** Band structure including fatbands of MgNiGe.

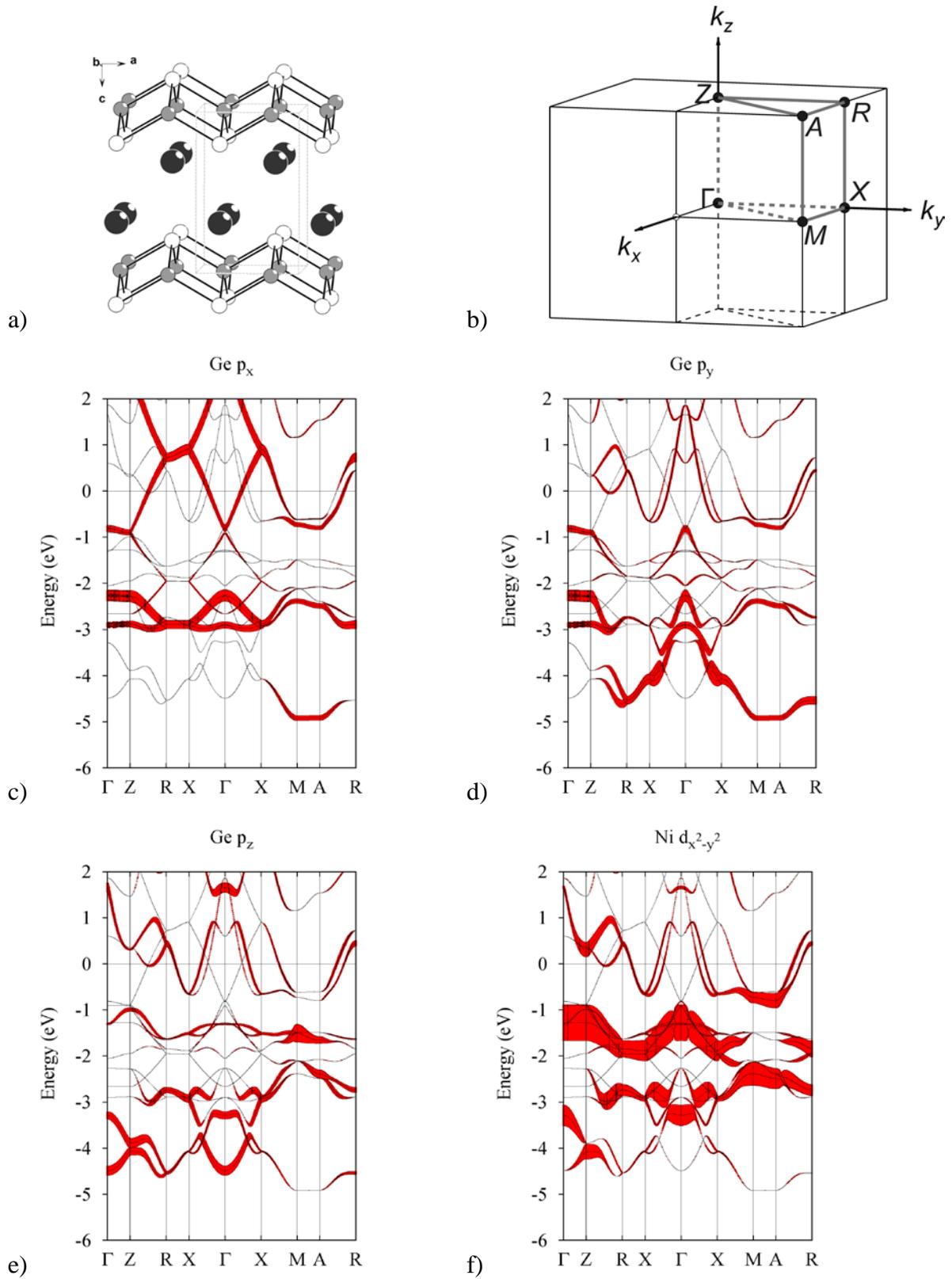

**Figure S9.** Band structure including fatbands of CaNiGe.

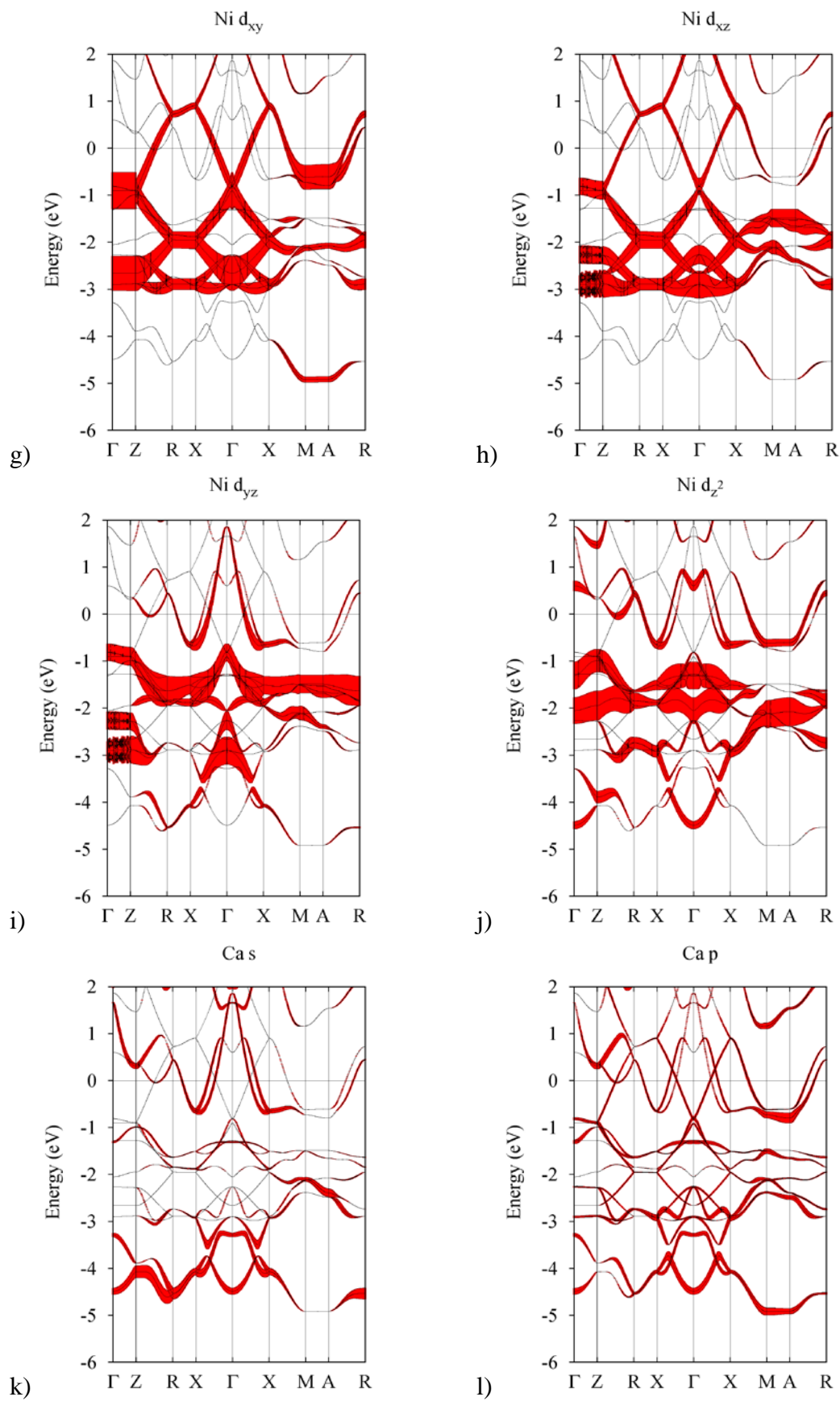

**Figure S9 (continued).** Band structure including fatbands of CaNiGe.

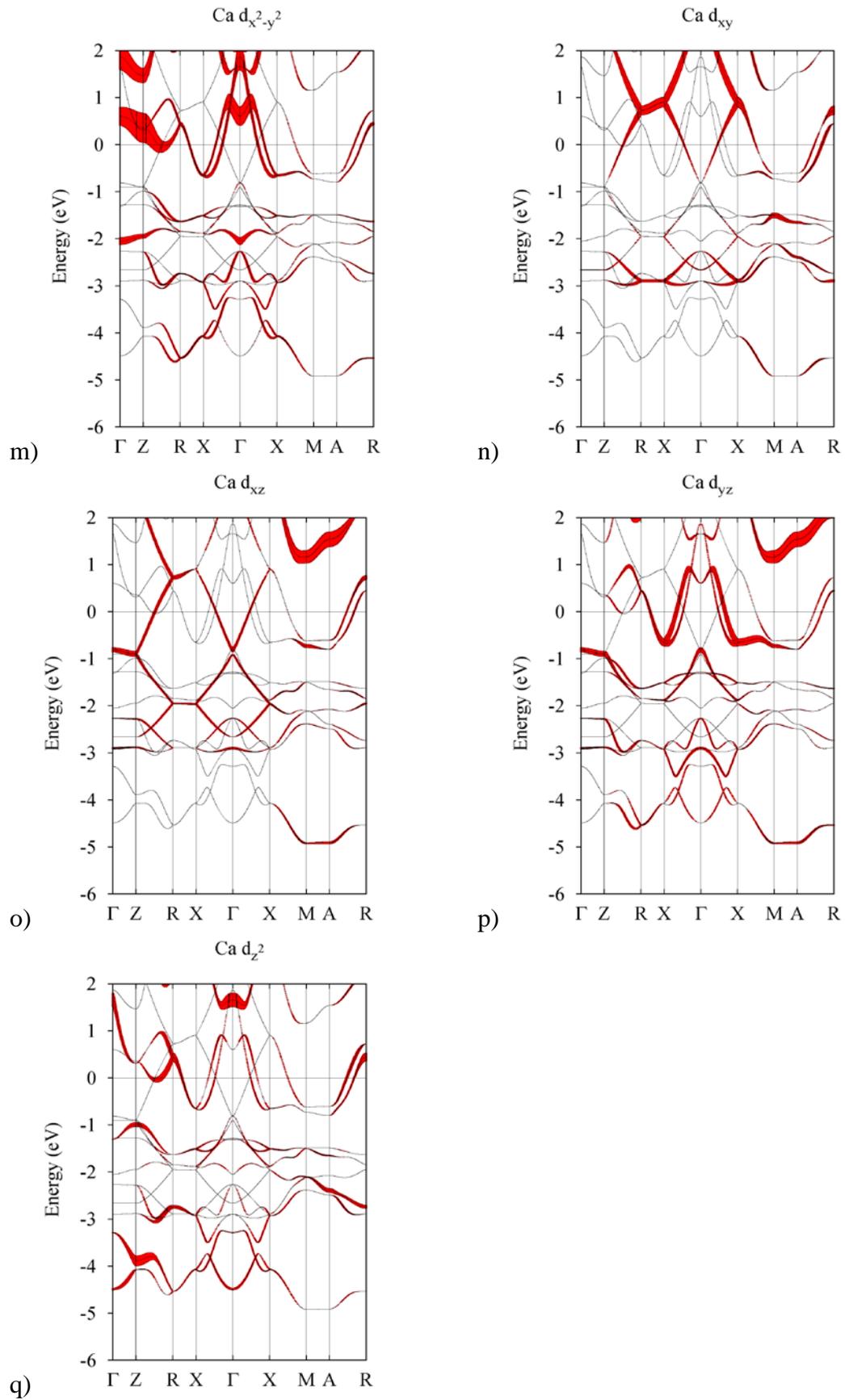

**Figure S9 (continued).** Band structure including fatbands of CaNiGe.

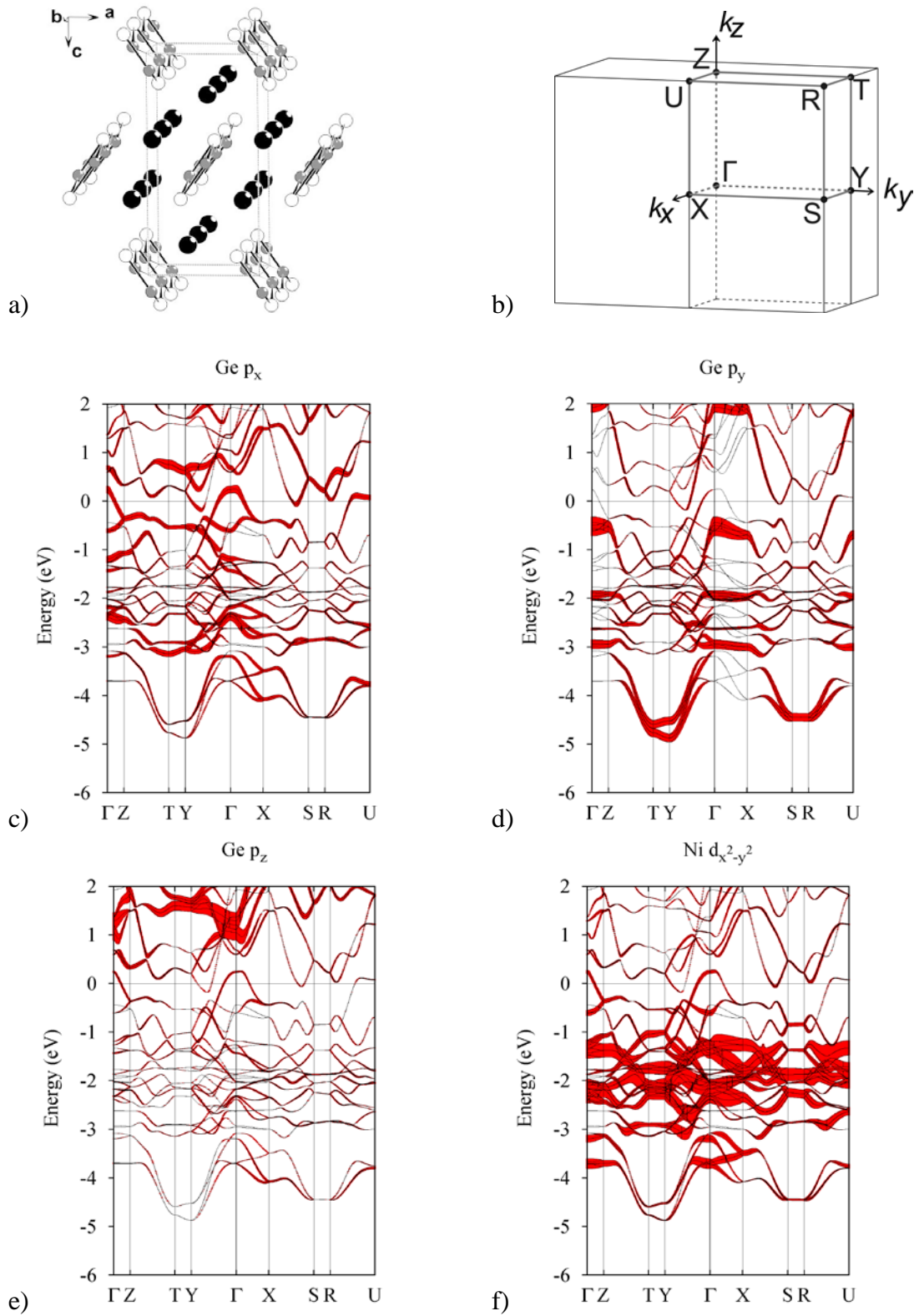

**Figure S10.** Band structure including fatbands of SrNiGe.

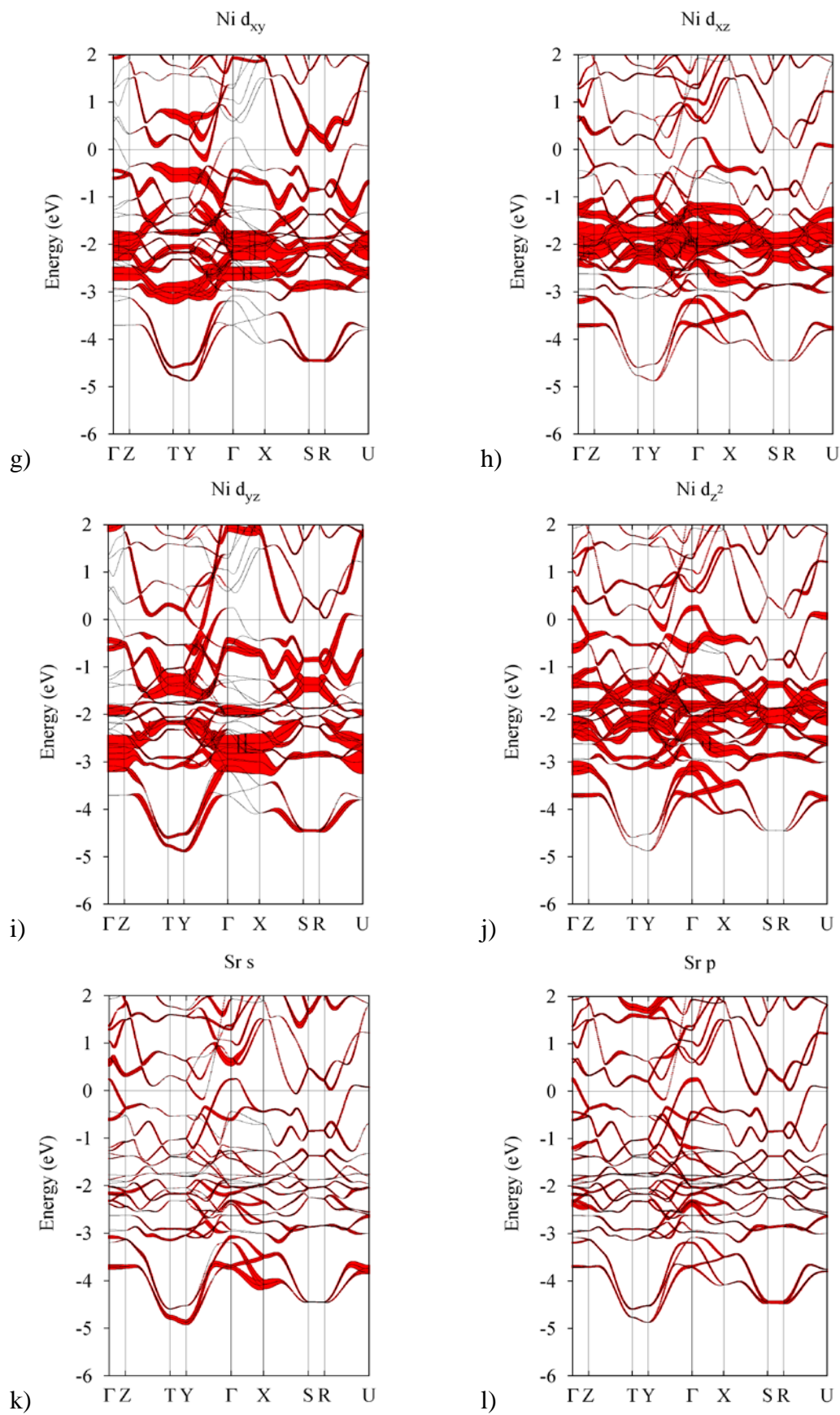

**Figure S10 (continued).** Band structure including fatbands of SrNiGe.

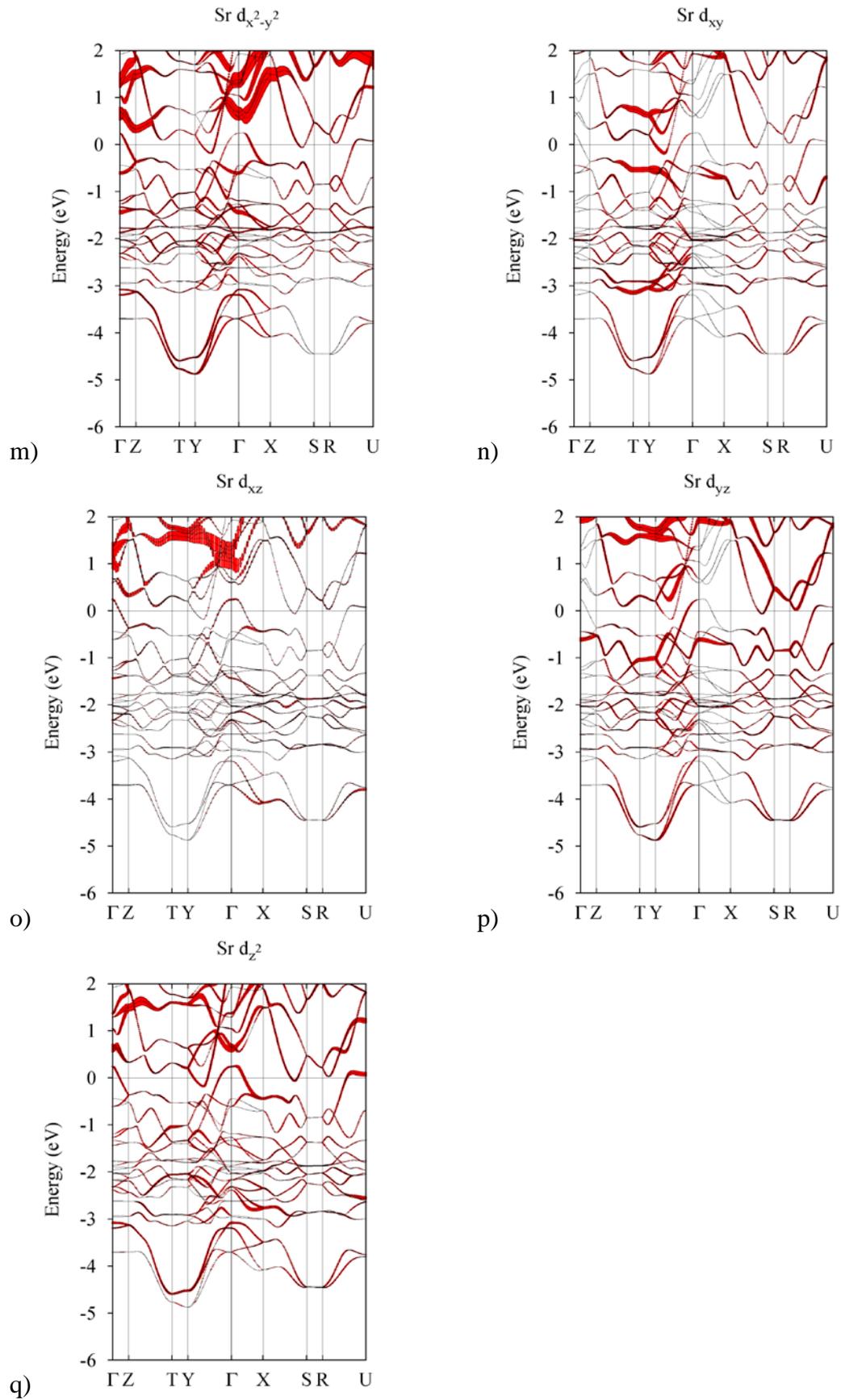

**Figure S10 (continued).** Band structure including fatbands of SrNiGe.

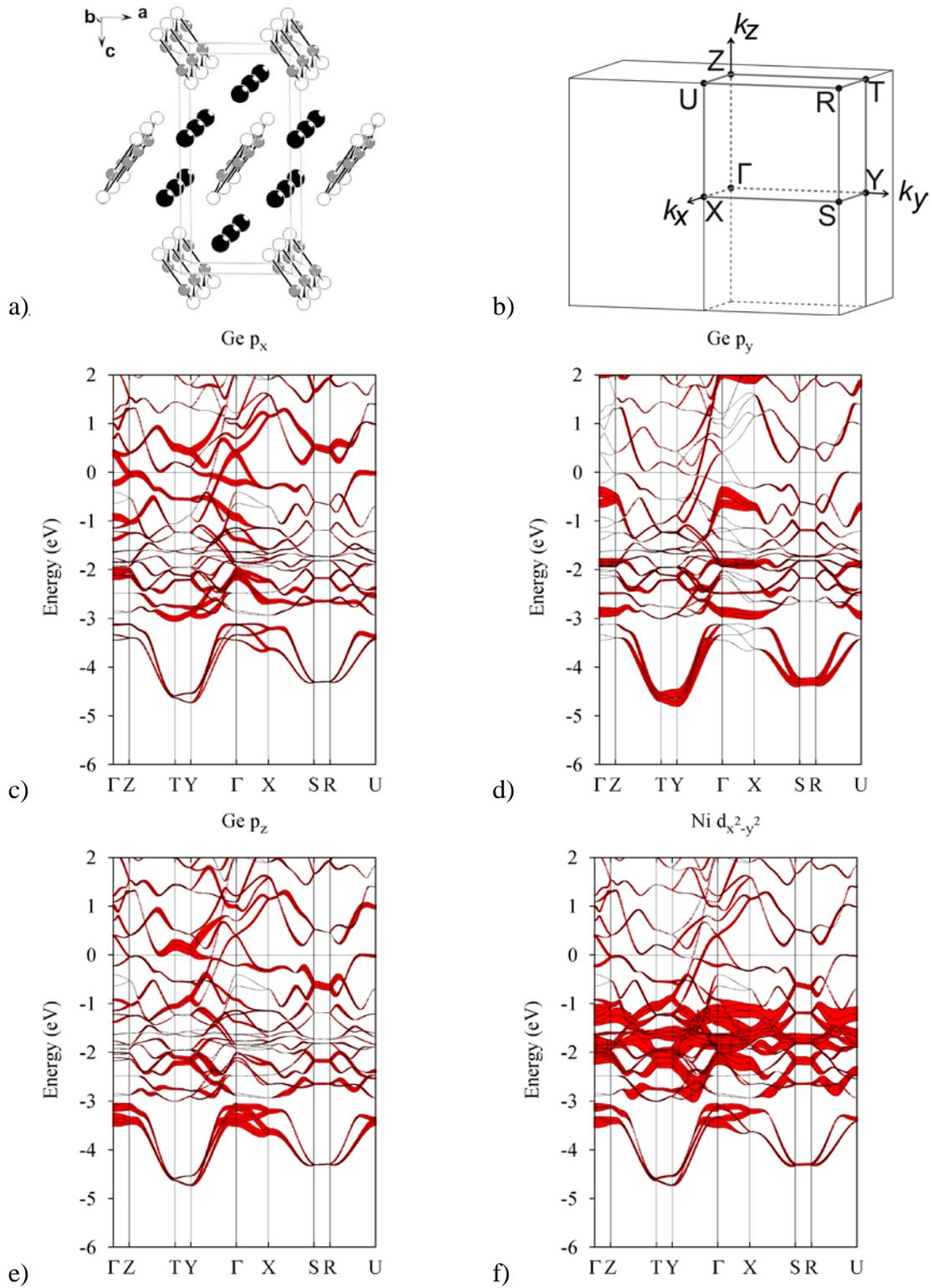

**Figure S11.** Band structure including fatbands of BaNiGe.

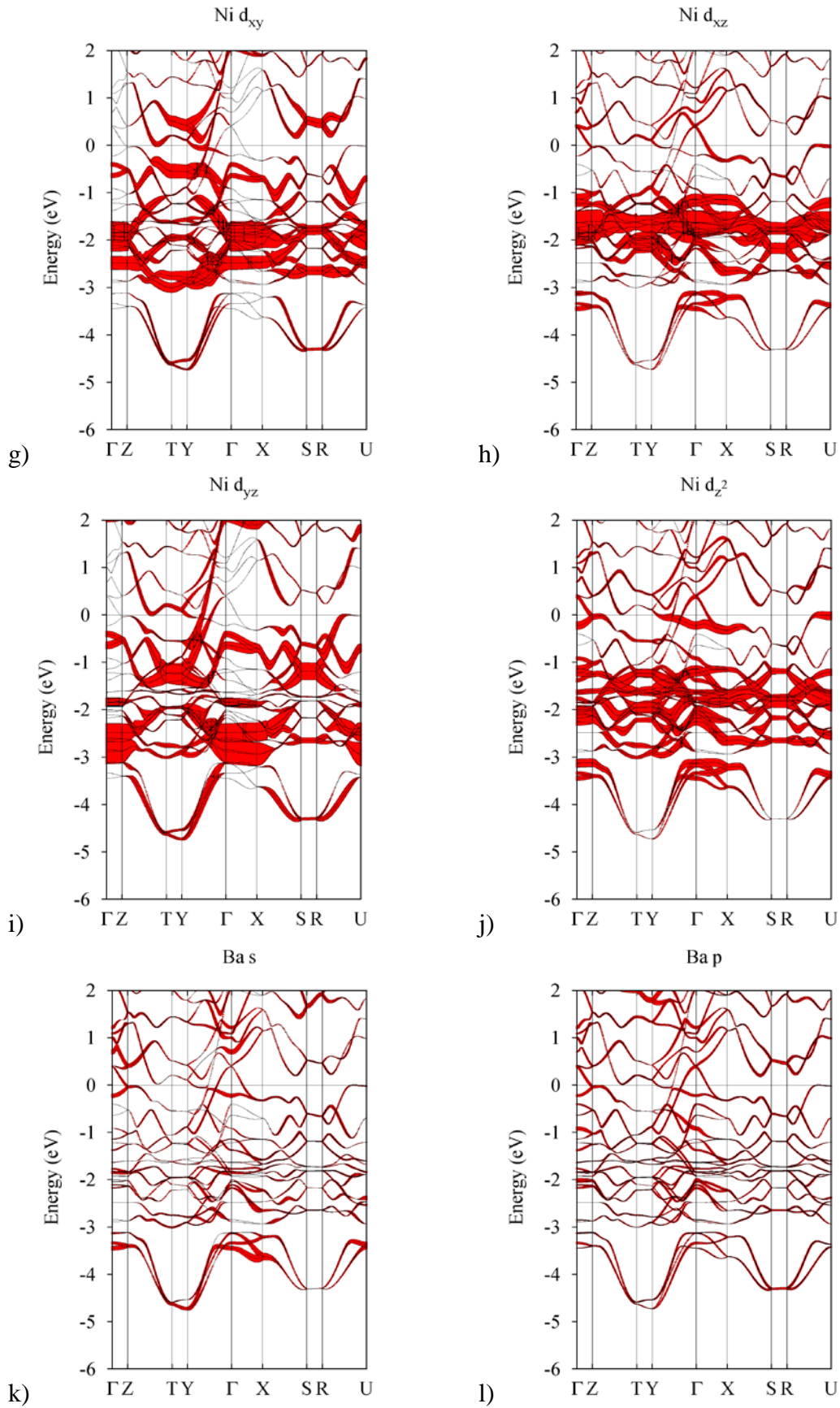

**Figure S11 (continued).** Band structure including fatbands of BaNiGe.

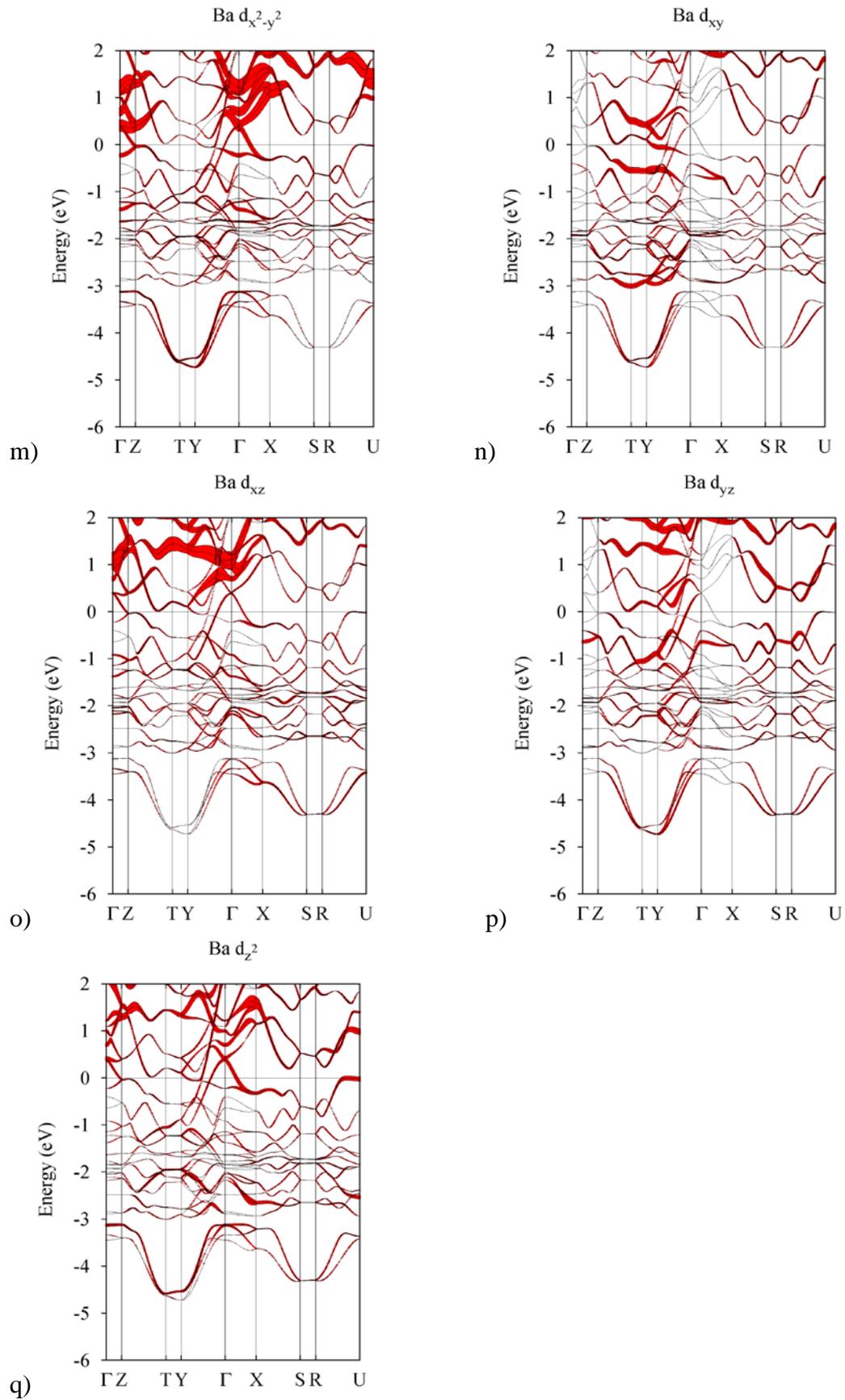

**Figure S11 (continued).** Band structure including fatbands of BaNiGe.

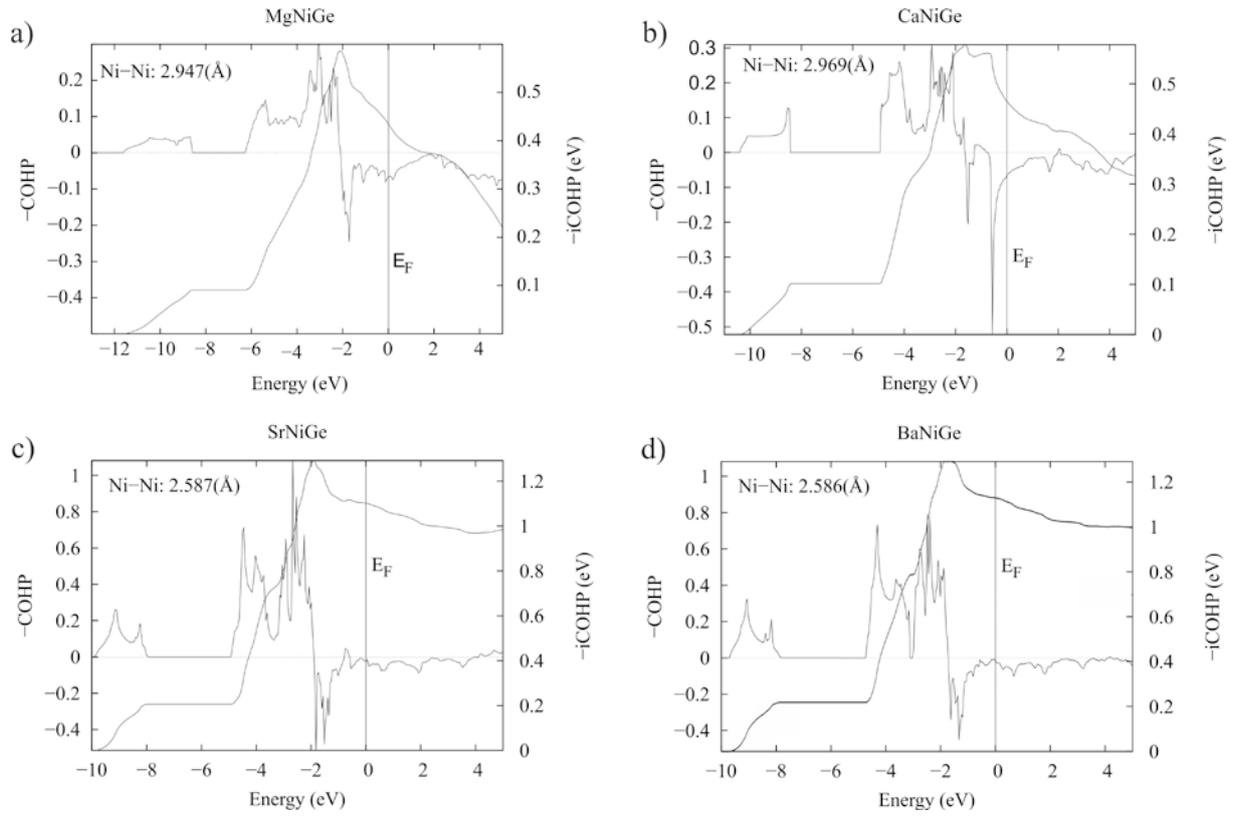

**Figure S12.** COHP of the Ni-Ni interaction for a) MgNiGe, b) CaNiGe, c) SrNiGe, d) BaNiGe.

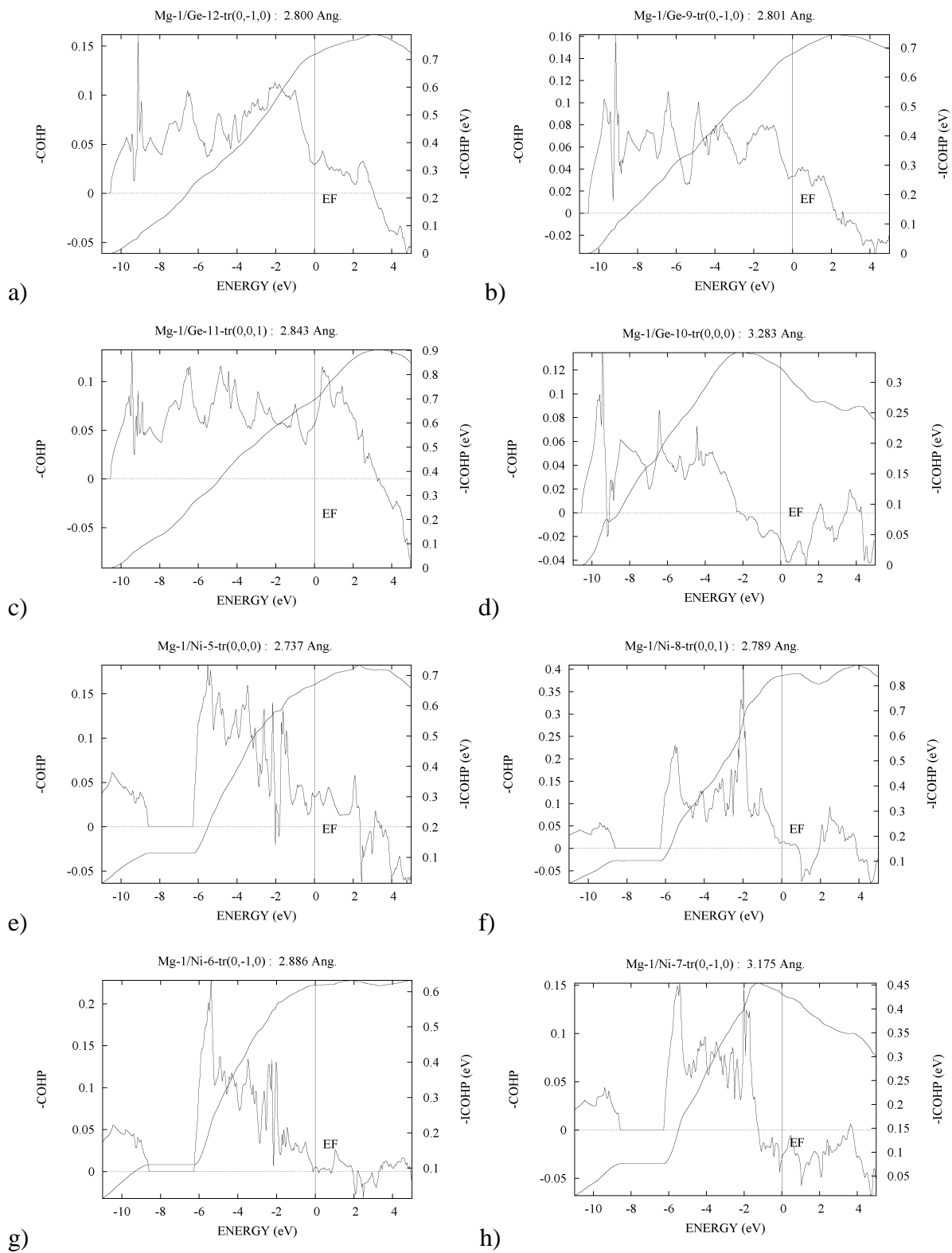

**Figure S13.** Crystal Orbital Hamilton Populations (COHP) and integrated crystal orbital Hamilton populations (−iCOHP) curves for corresponding Mg-Ge (a-d) and Mg-Ni (e-h) bonds for MgNiGe.

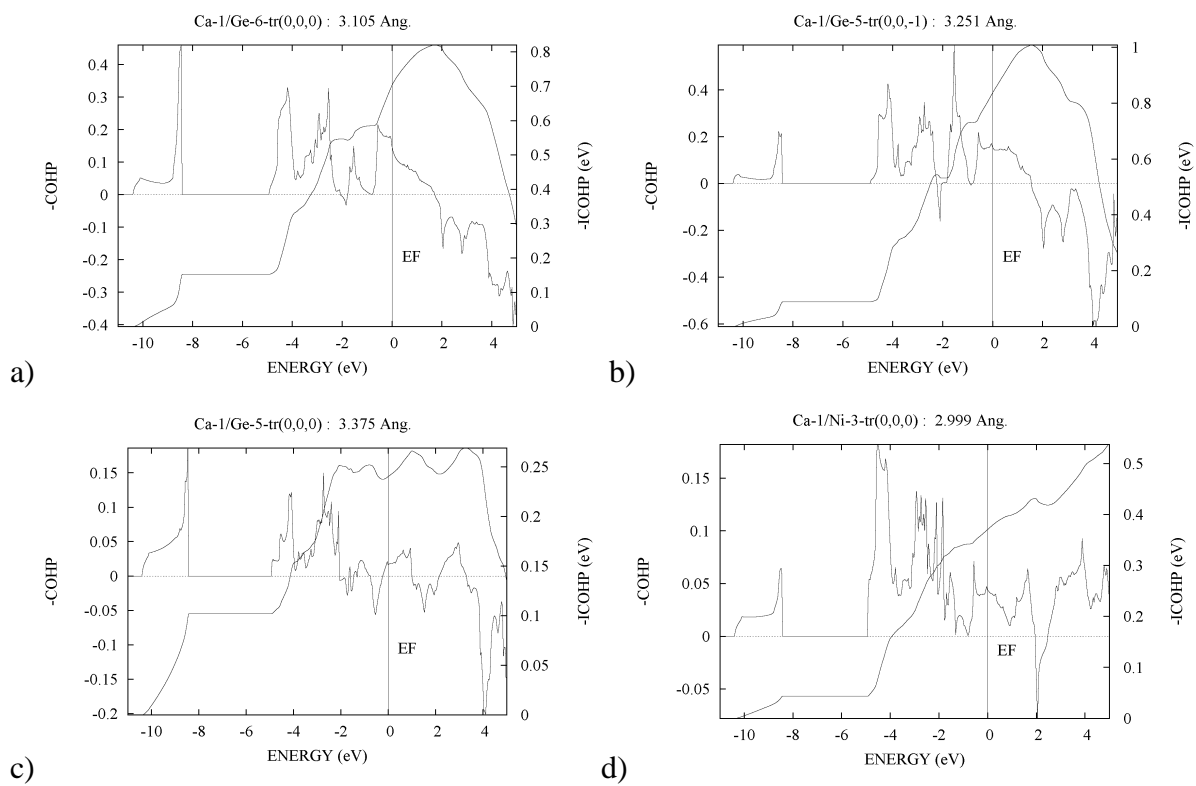

**Figure S14.** Crystal Orbital Hamilton Populations (COHP) and integrated crystal orbital Hamilton populations (−iCOHP) curves for corresponding Ca-Ge (a-c) and Ca-Ni (d) bonds for CaNiGe.

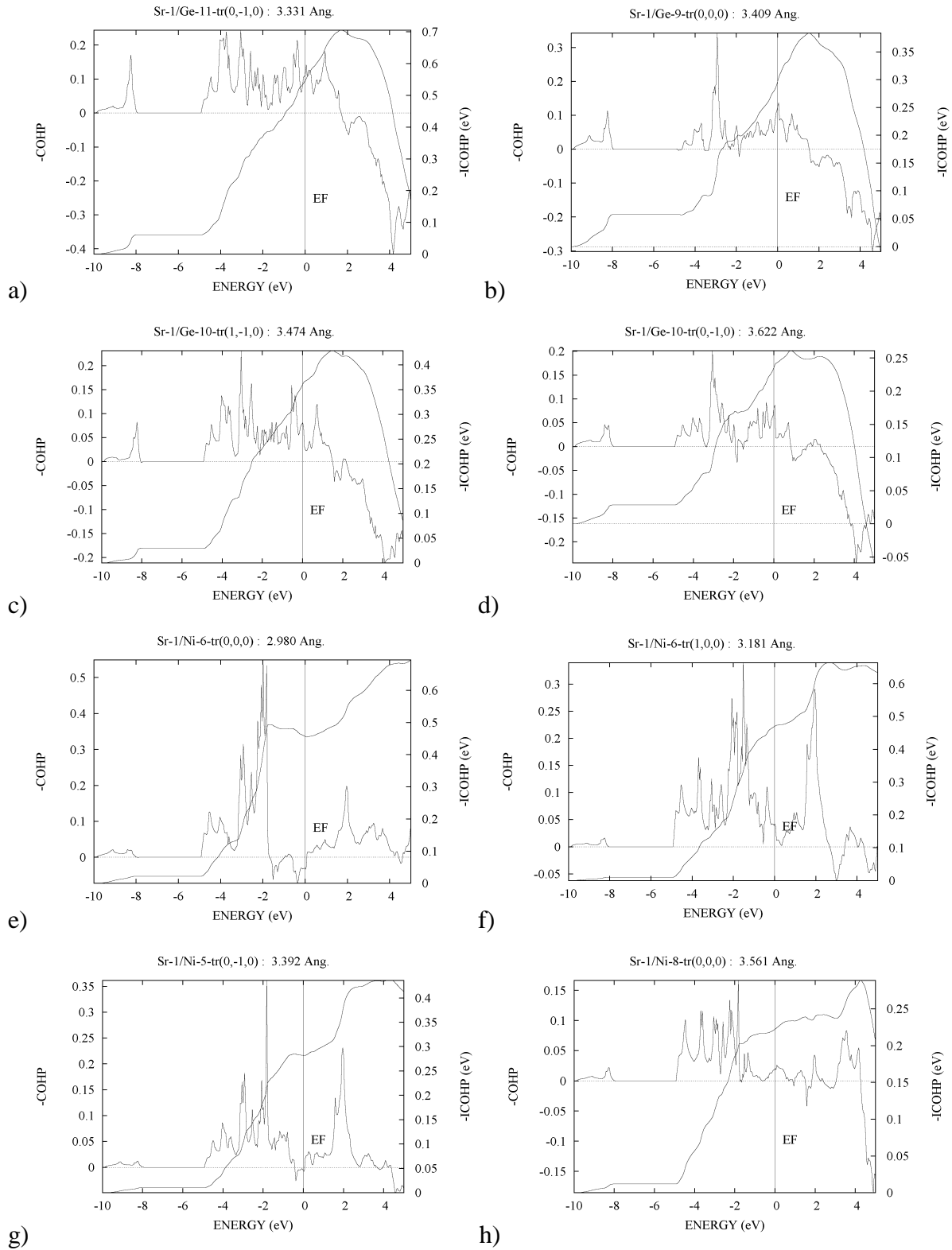

**Figure S15.** Crystal Orbital Hamilton Populations (COHP) and integrated crystal orbital Hamilton populations (−iCOHP) curves for corresponding Sr-Ge (a-d) and Sr-Ni (e-h) bonds for SrNiGe.

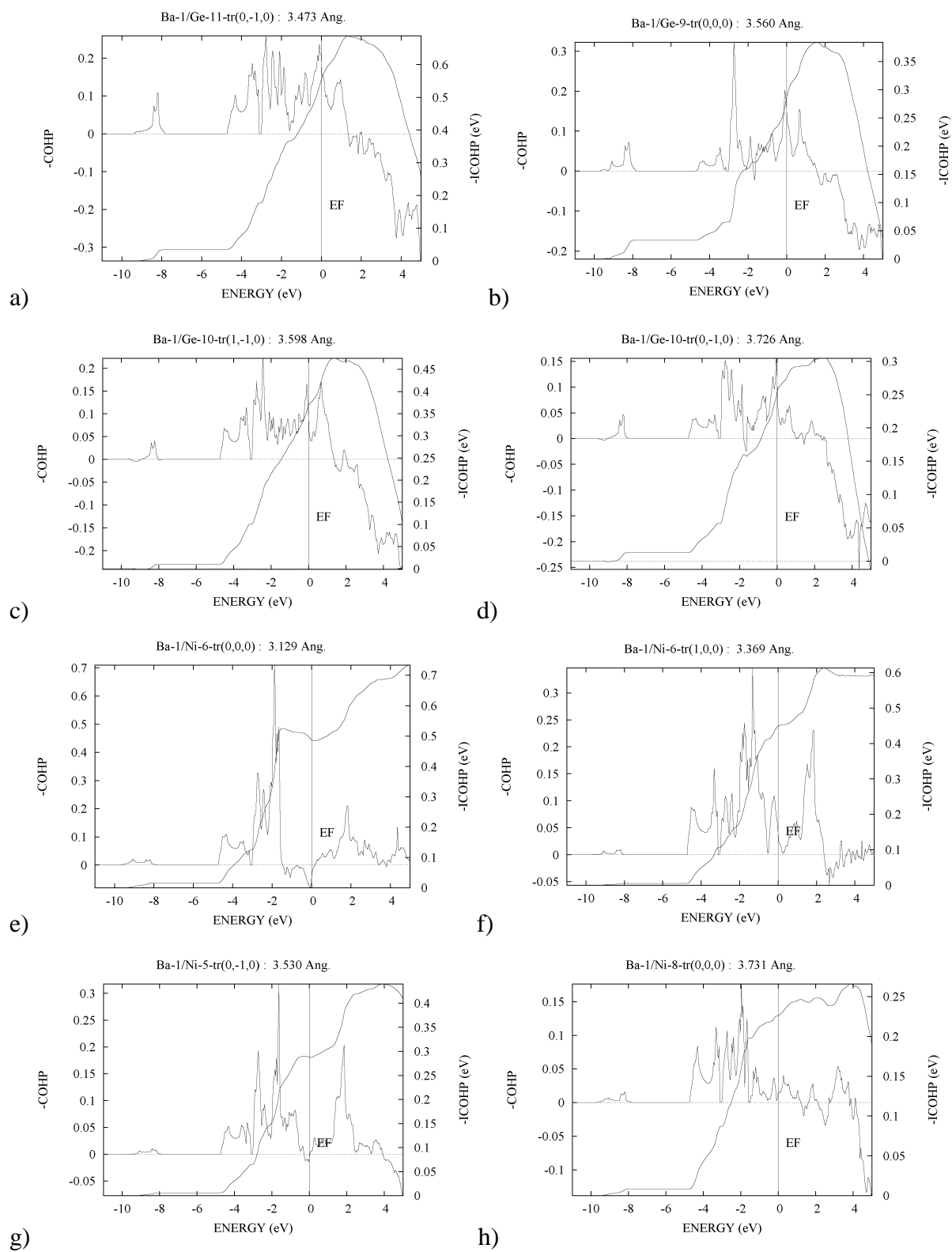

**Figure S16.** Crystal Orbital Hamilton Populations (COHP) and integrated crystal orbital Hamilton populations (−iCOHP) curves for corresponding Ba-Ge (a-d) and Ba-Ni (e-h) bonds for BaNiGe.